\documentclass[journal,twoside,web]{ieeecolor}

    \usepackage[pdftex]{graphicx}

\usepackage{generic}
\usepackage{gensymb}
\usepackage{epstopdf}
\usepackage{graphicx}
\usepackage{amssymb}
\usepackage{amsmath}
\usepackage{float}
\usepackage{amsfonts, bm}
\usepackage{color}
\usepackage[noadjust]{cite}

\DeclareMathOperator{\Ex}{\mathcal{E}}

\renewcommand{\theenumi}{\alph{enumi}}

\newtheorem{th1}{Theorem}

\title{\LARGE\textbf{Multi-Objective Vector Control of a Three-Phase  \\ Vibratory Energy Harvester}%
}

\author{
Connor H. Ligeikis and Jeffrey T. Scruggs
\thanks{The first author was supported by an NSF Graduate Research Fellowship. This funding is gratefully acknowledged. Views expressed in this paper are those of the authors and do not necessarily reflect those of the National Science Foundation.}
\thanks{C. Ligeikis and J. Scruggs are with the Department of Civil \& Environmental Engineering,
      	University of Michigan, Ann Arbor, MI, 48109.
      	Phone: \mbox{734-764-1812}, email: \mbox{ligeikis@umich.edu}, \mbox{jscruggs@umich.edu}}
}

\begin{document} 

\maketitle

\begin{abstract}

In vibration energy harvesting technologies, feedback control is required to maximize the average power generated from stochastic disturbances. 
In large-scale applications it is often advantageous to use three-phase conversion technologies for transduction. 
In such situations, vector control techniques can be used to optimally control the transducer currents in the direct-quadrature reference frame, as dynamic functions of feedback measurements.
In this paradigm, converted energy is optimally controlled via the quadrature current.
The direct current is only used to maintain control of the quadrature current when the machine's internal back-EMF exceeds the voltage of the power bus, a technique called field weakening. 
Due to increased dissipation in the stator coil, the use of field weakening results in a reduction in power conversion, relative to what would theoretically be possible with a larger bus voltage.
This over-voltage issue can be alternatively addressed by imposing a competing objective in the optimization of the quadrature current controller, such that the frequency and duration of these over-voltage events are reduced. 
However, this also results in reduced generated power, due to the need to satisfy the competing constraint.
This paper examines the tradeoff between these two approaches to over-voltage compensation, and illustrates a methodology for determining the optimum balance between the two approaches.

\vspace{6pt}

\end{abstract}



\section{Introduction}

Over the last two decades, an immense amount of research has been conducted on technologies to harvest energy from mechanical vibrations. 
The majority of this work has focused on small-scale technologies, intended for power levels below one milliwatt and frequencies above about 25Hz \cite{wei2017comprehensive}.  
For such applications, several modes of transduction have been successfully demonstrated, including piezoelectric \cite{erturk2011piezoelectric}, electromagnetic \cite{beeby2009electromagnetic}, and electrostatic \cite{khan2016state} technologies, as well as others.
Typically, the transducer is embedded within a resonant mechanical assembly, which is tuned to resonate at the dominant excitation frequency of the vibration energy to be harvested.
This assembly is then dynamically coupled to the vibratory phenomenon, and the transducer is interfaced with an isolated power bus or rechargeable storage system, thus facilitating a path for energy conversion.
Such energy harvesters enable sensing and computational technologies to be operated in energy-autonomy.
For example, they can be used to power sensors embedded within civil structures, which vibrate when subjected to vehicular and pedestrian traffic loads, as well as to various machinery \cite{cahill2018vibration,rhimi2012tunable,erturk2011piezoelectric,khan2016review}.
Although the intensity of this ambient vibration is often very low, it may provide sufficient energy to enable a wireless sensor to briefly power itself on once a day, take a measurement, transmit this measurement to a server, and power off again.

In theory, feedback control can be used to optimize the average power generated from stochastic vibrations \cite{scruggs2009optimal}. 
However, in the low-power and high-frequency regime, implementation of such feedback laws is not practical. 
This is because the theoretically-optimal feedback law typically requires that the transducer current be controlled continuously using high-frequency pulse-width-modulation (PWM). 
However, the level of available mechanical power may be so low that it is less than that which is necessary to switch a single MOSFET in PWM, due to the parasitics consumed by the gate drive circuit \cite{ottman2003optimized,liu2009active}. Consequently, more favorable performance can therefore be achieved in practice, with power-electronic circuits that switch at only very low frequencies (such as synchronized switching and related circuit topologies \cite{guyomar2005toward,guyomar2011recent}), or circuits that involve no controllable switching at all (such as a simple diode bridge rectifier).  Although such circuits are theoretically sub-optimal under the assumption of zero parasitic loss, their performance can be superior to circuits requiring PWM switching when these losses are taken into account.

\begin{figure*}
\centering
   \includegraphics[scale=0.43]{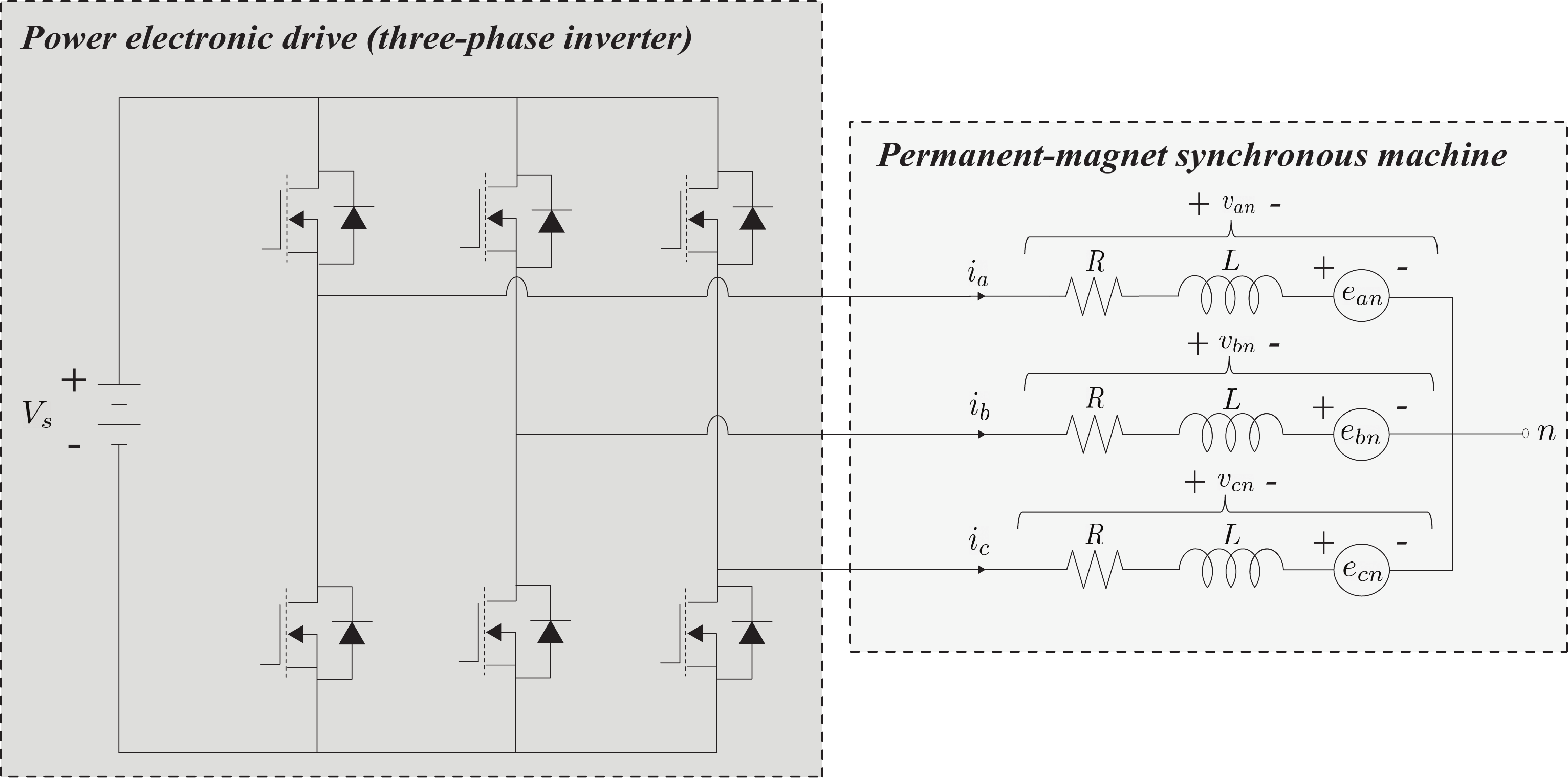}
\caption{Standard power-electronic drive interfaced with three-phase permanent-magnet synchronous machine}
\vspace{-5pt}
\label{inverter}
\end{figure*}

Vibration energy harvesting is also useful at larger power scales, and at lower frequencies. 
Arguably the most important application in this regime concerns the generation of utility-scale power from ocean waves. 
Wave energy conversion (WEC) technologies are emerging as a promising alternative to wind, solar, and geothermal sources of renewable energy \cite{astariz2015economics}. 
WECs often operate at average power levels in excess of 50kW, and at frequencies between $0.05-0.2$Hz. 
As another example, energy harvesting technology can be used to capture power from the vibratory responses of wind-excited buildings, at power scales above $1$kW and frequencies below $1$Hz \cite{zuo2013large}. 
The harvested energy in these applications can, in turn, be used to power the feedback control systems that optimize vibration suppression, resulting in closed-loop systems that operate in energy-autonomy \cite{ligeikis2021nonlinear,li2022self}. 
Similar self-powered control technologies can be implemented in high-performance vehicle suspensions, for the purpose of minimizing cabin accelerations \cite{nakano2003self,abdelkareem2018vibration}. 
Typical power levels in such applications are on the order of $10$W, with vibratory frequencies of about $1$Hz.

For these larger-scale energy harvesting technologies, the time-averaged power available for generation far exceeds parasitic power dissipation, even when PWM-controlled power electronics are used to continuously regulate transducer currents.
Consequently, optimal feedback control theory can justifiably be used to maximize average power generation.
It has been shown that if the vibratory disturbance is stationary stochastic, the dynamically-excited plant is linear, and if the primary parasitic conductive losses are quadratic (i.e., $I^2R$) losses, then the optimal energy-harvesting feedback law is the solution to a sign-indefinite Linear Quadratic Gaussian (LQG) control problem \cite{Scruggs2010}.
Numerous studies have been conducted on the use of various related optimal control techniques, which can accommodate nonlinearities in the harvester dynamics \cite{nie2016optimal}, non-quadratic loss models \cite{Scruggs2012}, and non-stationary disturbances \cite{kody2019control}.
Furthermore, some recent work (related to WEC systems) has been done on the use of adaptive control techniques to autonomously accommodate changes in the spectrum of the vibratory disturbance, as well as variability in the plant dynamics \cite{scruggs2015disturbance,davidson2018adaptive}. 
Beyond these studies, a vast amount of work has been published on the use of Model-Predictive Control (MPC) to optimize performance of large-scale vibratory energy harvesters, for the case in which future disturbances are either known or can be accurately forecast \cite{richter2012nonlinear,li2014model,faedo2017optimal}.  
(In WEC applications particularly, such disturbance forecasts may be practical \cite{ringwood2014energy}.)
Such MPC techniques have the advantage of straight-forwardly accommodating constraints, both for control inputs as well as for response quantities. 

At larger power scales, three-phase, permanent-magnet synchronous machines (PMSMs) are often used as transducers \cite{scruggs2018analysis}.
The use of three-phase machines is preferable, compared to DC machines, because they typically have much higher power density, they are more efficient, and they are commercially-available in higher power ratings. 
The use of PMSMs, specifically, is advantageous because they are efficient and power-dense, and they are easy to control over a wide dynamic operating regime.
Although direct-drive linear machines are sometimes used \cite{polinder2003linear,mueller2002electrical,danielsson2005direct}, rotary machines are common, especially when it is important to economize mass and size \cite{sabzehgar2013energy,maravandi2015regenerative,liu2017design,mccullagh2019analysis,pan2018compact,wang2020high,sugiura2020experimental}. 
In such cases, the rotor is interfaced with the rectilinear vibratory motion of the energy harvester through one of a variety of mechanical mechanisms, such as a rack-and-pinion, ballscrew, or planetary roller screw mechanism. 
When designed properly, such linear-to-rotational conversion mechanisms can achieve efficiencies in excess of 90\% in both forward-drive and backdrive operation.

The standard power-electronic drive that interfaces a three-phase PMSM with a DC power bus is illustrated in Figure~\ref{inverter}.
In PWM operation, the six MOSFETs are switched on and off at high frequency, so as to track desired phase currents $i_a$ and $i_b$, with $i_c = -i_a-i_b$. 
If the velocity of the machine is sufficiently large, the magnitudes of its line-to-line back-EMFs can exceed the bus voltage $V_s$, in which case the phase currents must be expressly controlled so as to counteract the field of the rotor, in order to maintain controllability of the drive \cite{soong2002field}.
This technique, called \emph{field weakening}, allows for a lower bus voltage to be used for operation in a given dynamic response regime, which can result in lower parasitic switching losses in the drive. 
However, the use of field weakening also leads to higher conductive losses in the stator coils.  
Consequently, for a given dynamic response regime, the choice of $V_s$ constitutes a trade-off, and in general there exists an optimal value that maximizes efficiency \cite{scruggs2018analysis}.

In the context of energy harvesting, the dynamic response regime is affected by the manner in which the stator currents are controlled.
There is an optimal causal feedback law, relating the dynamic output measurements of the harvester to the stator currents, which maximizes average generated power in stationary stochastic response. 
With this optimal feedback imposed, the vibratory intensity of the harvester velocity (and therefore the back-EMFs of the machine) is significant, and under high excitation, field-weakening may be necessary.
Alternatively, in order to maintain back-EMF response amplitudes at levels below the bus voltage $V_s$, a competing objective may be imposed on the optimization of the feedback law, which enforces a bound on the vibratory response intensity of the harvester velocity.
This multi-objective optimal control approach can be used as an alternative to field weakening, or in tandem with it, as a means of accommodating the finite bus voltage.  
However, as with field-weakening, this approach involves a compromise in power generation performance. 
This is because, by imposing a competing vibration-suppression objective in the optimization of the feedback controller, the primary energy-harvesting objective is reduced from its unconstrained optimum value. 

The motivation for this paper is to characterize the tradeoff between these two techniques (i.e., field weakening and velocity suppression) for accommodating a finite bus voltage in a vibration energy harvesting application. 
Further, we establish and experimentally validate a procedure to determine the optimal combination of the two techniques, to maximize power generation.
The specific contributions of the paper are as follows. 
In Section II we provide an overview of the nonlinear stochastic model for the dynamics of a vibration energy harvester with a three-phase PMSM as a transducer.  
In Section IIIA-C, we illustrate a systematic technique for multi-objective optimization of a dynamic output-feedback law for a stochastically-excited energy harvester with linear dynamics, such that the average power generation is maximized. 
In Section IIID-E, we illustrate the extension of this methodology to accommodate field weakening, given a finite bus voltage $V_s$. 
In Section IIIF-H, we illustrate the further extension of the methodology to accommodate nonlinearities in the harvester dynamics, using the principle of stochastic linearization, resulting in an iterative optimization procedure.
In Section IV, we use simulation to determine the optimal balance between the two means of accommodating a finite bus voltage (i.e., field weakening vs. vibration suppression).
In Section V, we validate the optimized control design experimentally, in the form of a Hardware-in-the-Loop (HiL) experiment.    
Finally, Section VI provides some conclusions. 

\section{Modeling}

\subsection{Mechanical dynamic model}

\begin{figure*}
    \vspace{0pt}
    \centering
   \includegraphics[scale=0.5,trim={0cm 0cm 0cm 0cm},clip]{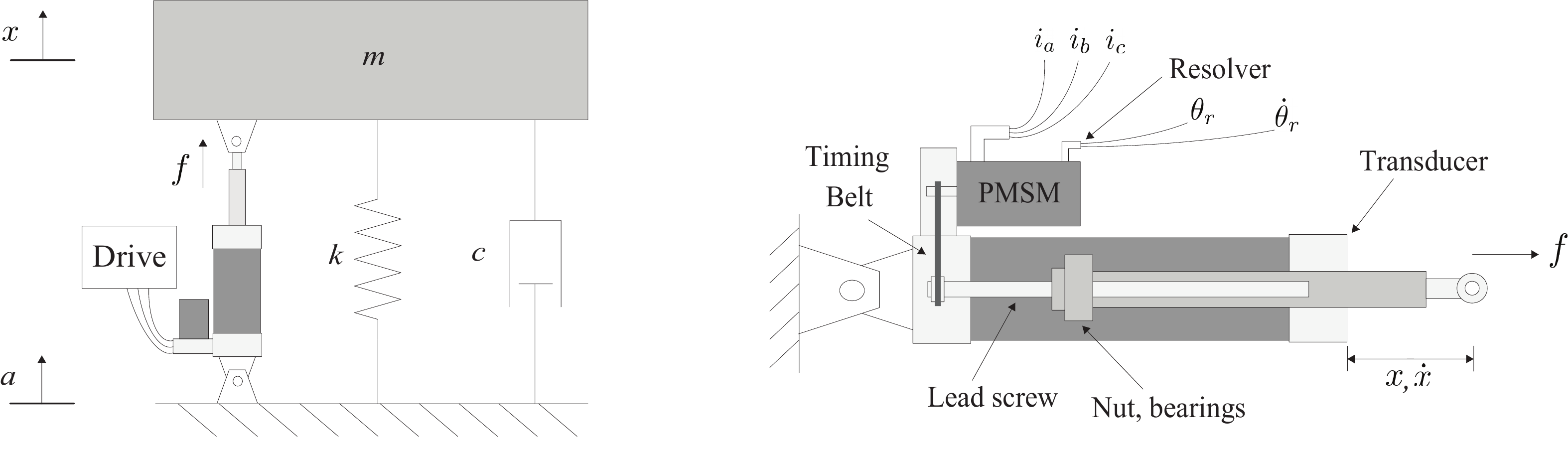}
   \caption{Single-degree-of-freedom energy harvester (left) and permanent-magnet synchronous machine (PMSM) transducer with internal components illustrated (right)}
   \label{fig_trans}
   \vspace{-5pt}
\end{figure*}

Consider the vibratory energy harvesting system shown in Figure \ref{fig_trans}. 
It consists of a single-degree-of-freedom (SDOF) oscillator coupled with an electromechanical transducer. The transducer consists of a surface-mount rotary PMSM interfaced with a precision ballscrew via a timing belt to accomplish linear-to-rotational motion conversion. 
The motion of the energy harvester evolves according the following differential equation
\begin{equation}\label{sdof_eom}
    m\ddot{x}(t)+c\dot{x}(t)+k x(t) = -ma(t) + f(t)
\end{equation}
where $m$ is the SDOF mass, $c$ is the viscous damping, $k$ is the stiffness, $x(t)$ is the relative displacement of the mass, $a(t)$ is the stochastic base acceleration, and $f(t)$ is the force exerted on the mass by the transducer. 
We assume that the linear-to-rotational conversion has negligible backlash, and that the timing belt is infinitely stiff, resulting in static linear relationship between $x(t)$ and the PMSM's mechanical rotation angle $\theta_r(t)$.
The resultant relationship between the respective linear and rotational velocities is $\dot{x}(t) = \ell \dot{\theta}_r(t)$, where $\ell$ is the effective screw lead length, which includes an amplification factor due to belt drive ratio. 

The transducer force $f(t)$ is a consequence of several interacting physical phenomena.
The rotor has a finite rotary inertia $J > 0$ and viscous damping $B > 0$, which contribute apparent rectilinear inertia and damping terms to $f(t)$. 
Additionally, the rectilinear sliding between the ballscrew nut and the guide produces a Coulomb friction force $f_c$. 
The electromechanical conversion of energy by the PMSM results in an apparent rectilinear force $f_e(t)$ at the ballscrew nut.
And finally, the sliding of the bearings between the nut and the screw produces an approximately-static linear-to-rotational conversion efficiency $\eta \in (0,1)$. 
In  \cite{CassidySPIE2011,cassidy2012statistically} it is shown that these effects can be approximately modeled as
\begin{equation}\label{trans_model}
   f(t) = h(p(t)) \left(f_e(t) - \frac{J}{\ell^2}\ddot{x}(t) - \frac{B}{\ell^2}\dot{x}(t)\right) - f_c \textrm{sgn}(\dot{x}(t))
\end{equation}
where $p(t)$ is the mechanical power delivered to the nut, i.e.,
\begin{equation}
    p(t) = \left(f_e(t) - \frac{J}{\ell^2}\ddot{x}(t) - \frac{B}{\ell^2}\dot{x}(t)\right) \dot{x}(t) , 
\end{equation}
and where $h(\cdot)$ and $\textrm{sgn}(\cdot)$ are discontinuous functions, which satisfy
\begin{align}
    h(p(t)) & \begin{cases}
          = \eta &: \, p(t) > 0 \\
          \in [\eta,1/\eta] &: \, p(t) = 0 \\
          = 1/{\eta} &: \,  p(t) < 0 
     \end{cases}
     \\
     \textrm{sgn}(\dot{x}(t)) & \begin{cases}
        = 1 &: \, \dot{x}(t) > 0 \\
        \in [-1,1] &: \, \dot{x}(t) = 0 \\
        = -1 &: \, \dot{x}(t) < 0 
     \end{cases}
\end{align}
Equation \eqref{trans_model} is imprecise, because $h(\cdot)$ and $\textrm{sgn}(\cdot)$ are not uniquely defined for the case in which their arguments are zero. 
However, when \eqref{trans_model} and \eqref{sdof_eom} are combined, the value of $f(t)$ is a unique static function of $\{x(t),\dot{x}(t),f_e(t),a(t)\}$. 

To show this, first consider the case in which $\dot{x}(t) > 0$.
In this case, it follows that 
\begin{align}
    \frac{J}{\ell^2m} f(t) \dot{x}(t) 
    =& \frac{J}{\ell^2m} h(p(t)) p(t) - \frac{J}{\ell^2m} f_c \dot{x} \\
    =& -p(t) + \beta \left( x(t), \dot{x}(t), f_e(t), a(t) \right) \dot{x}(t) 
\end{align}
where 
\begin{multline}
    \beta\left( x(t), \dot{x}(t), f_e(t), a(t) \right)
    \triangleq \frac{J}{\ell^2} a(t) + \frac{Jk}{m\ell^2} x(t)  \\ + \left[ \frac{Jc}{m\ell^2} - \frac{B}{\ell^2} \right] \dot{x}(t) + f_e(t).
\end{multline}
It follows that 
\begin{multline}
    \left[ 1 + \frac{J}{\ell^2m} h(p(t)) \right] p(t) = \\ \left[ \beta\left( x(t), \dot{x}(t), f_e(t), a(t) \right) + \frac{J}{\ell^2m} f_c \right] \dot{x}(t)
\end{multline}
The left-hand side is an invertible function of $p(t)$, and consequently the equation returns a unique solution for $p(t)$, for all $\{ x(t), \dot{x}(t), f_e(t), a(t) \} \in \mathbb{R}^4$.
For this solution, denote
\begin{align}
   &\Phi_+\left( x(t), \dot{x}(t), f_e(t), a(t) \right) \nonumber \\ 
   &\quad \triangleq \frac{ h(p(t)) p(t) }{\dot{x}(t)} - f_c \\
   &\quad = g_+\left( \beta\left( x(t), \dot{x}(t), f_e(t), a(t) \right) + \frac{J}{\ell^2m} f_c \right) - f_c 
\end{align}
where 
\begin{equation}
    g_+(u) \triangleq \frac{ h(u) u }{ 1 + \frac{J}{\ell^2m} h(u) }
\end{equation}
is continuous for all $u \in \mathbb{R}$. 
Then we have that $f(t)$ can be found uniquely as
\begin{equation} \label{f_xdot_xdot>0}
    f(t) = \Phi_+\left( x(t), \dot{x}(t), f_e(t), a(t) \right) 
\end{equation}
An analogous process for the case in which $\dot{x}(t) < 0$ gives that
\begin{equation} \label{f_xdot_xdot<0}
    f(t) = \Phi_{-}\left( x(t), \dot{x}(t), f_e(t), a(t) \right) 
\end{equation}
where
\begin{align}
   &\Phi_-\left( x(t), \dot{x}(t), f_e(t), a(t) \right) \nonumber \\ 
   &\quad \triangleq \frac{ h(p(t)) p(t) }{\dot{x}(t)} + f_c \\
   &\quad = g_-\left( \beta\left( x(t), \dot{x}(t), f_e(t), a(t) \right) - \frac{J}{\ell^2m} f_c \right) + f_c 
\end{align}
and where 
\begin{equation}
    g_-(u) \triangleq \frac{ h(-u) u }{ 1 + \frac{J}{\ell^2m} h(-u) }
\end{equation}
is continuous for all $u \in \mathbb{R}$. 
For $\{x(t),\dot{x}(t),f_e(t),a(t)\} \in \mathbb{R}^4$ with $\dot{x}(t) = 0$, the Coulomb friction force constrains the trajectory to slide on this subspace (i.e., enforces $\ddot{x}(t) = 0$) if the magnitude of the force required to do so is less than $f_c$. 
It is straight-forward to show that this is the case if and only if
\begin{multline} \label{fc_ineq_0}
    f_c \geqslant \max\big\{ -ma(t)-kx(t) + h(f_e(t)) f_e(t) , \\ ma(t)+kx(t)-h(-f_e(t)) f_e(t) \big\} 
\end{multline}
in which case it follows that 
\begin{equation} \label{f_xdot_zero_0}
    f(t) = ma(t) + kx(t)
\end{equation}
Otherwise, if
\begin{equation} \label{fc_ineq_1}
    f_c < -ma(t)-kx(t) + h(f_e(t)) f_e(t)
\end{equation}
then the friction force is equal to its lower bound, and
\begin{align} \label{f_xdot_zero_1}
    f(t) =& h(f_e(t)) \left( f_e(t) - \frac{J}{\ell^2} \ddot{x}(t) \right) - f_c  \\
    =& \Phi_+ \left( x(t), 0, f_e(t), a(t) \right)
\end{align}
resulting in $\ddot{x}(t) > 0$.
Likewise, if 
\begin{equation} \label{fc_ineq_2}
    f_c < ma(t)+kx(t) - h(-f_e(t)) f_e(t)
\end{equation}
then the friction force is equal to its upper bound, and
\begin{align} \label{f_xdot_zero_2}
    f(t) =& h(-f_e(t)) \left( f_e(t) - \frac{J}{\ell^2} \ddot{x}(t) \right) + f_c \\
    =& \Phi_-\left( x(t), 0, f_e(t), a(t) \right)
\end{align}
resulting in $\ddot{x}(t) < 0$. 
We note that both \eqref{fc_ineq_1} and \eqref{fc_ineq_2} cannot simultaneously be true because the sums of the right-hand sides of these inequalities is nonpositive for all $f_e(t) \in \mathbb{R}$ and all $\eta \in (0,1)$. 

To summarize, we have the unique mapping $\{x(t),\dot{x}(t),f_e(t),a(t)\} \mapsto f(t)$, as
\begin{align}
    f(t) =& \left\{ \begin{array}{l}
        \Phi_+\left( x(t), \dot{x}(t), f_e(t), a(t) \right) \\
        \quad\quad :\, \{ x(t),\dot{x}(t),f_e(t),a(t)\} \in \mathbb{S}_+ \\
        \Phi_-\left( x(t), \dot{x}(t), f_e(t), a(t) \right) \\
        \quad\quad :\, \{ x(t),\dot{x}(t),f_e(t),a(t)\} \in \mathbb{S}_- \\
        ma(t) + kx(t) \\
        \quad\quad :\, \{ x(t),\dot{x}(t),f_e(t),a(t)\} \notin \mathbb{S}_-\cup\mathbb{S}_+
    \end{array} \right.
    \\ \label{f_complete}
    \triangleq & \Phi \left(  x(t), \dot{x}(t), f_e(t), a(t) \right) 
\end{align}
where sets $\mathbb{S}_+$ and $\mathbb{S}_-$ are
\begin{align}
    \mathbb{S}_+ =& \big\{ x(t),\dot{x}(t),f_e(t),a(t) \ : \ \dot{x}(t) > 0  \nonumber \\ & \quad\quad\quad \quad\quad\quad 
    \lor \ \left( \dot{x}(t) = 0 \ \land \ \eqref{fc_ineq_1} \right) \big\} \\
    \mathbb{S}_- =& \big\{ x(t),\dot{x}(t),f_e(t),a(t) \ : \ \dot{x}(t) < 0  \nonumber \\ & \quad\quad\quad \quad\quad\quad 
    \lor \ \left( \dot{x}(t) = 0 \ \land \ \eqref{fc_ineq_2} \right) \big\}
\end{align}

\subsection{Electrical dynamic model}

Electromechanical force $f_e(t)$ is determined by the PMSM's three-phase currents, which evolve according to
\begin{equation} \label{stat_eqn}
    \frac{d}{dt}i_{abc}(t) = \frac{1}{L} \left( v_{abc}(t) - R i_{abc}(t) + e_{abc}(t) \right)
\end{equation}
where $i_{abc}(t)\triangleq\begin{bmatrix} i_a(t) & i_b(t) & i_c(t) \end{bmatrix}^T$ is the vector of three-phase line-to-neutral currents, $v_{abc}(t)\triangleq\begin{bmatrix} v_{an}(t) & v_{bn}(t) & v_{cn}(t) \end{bmatrix}^T$ is the vector of three-phase line-to-neutral stator voltages, $L$ is the line-to-neutral winding inductance, $R$ is the line-to-neutral winding resistance, and $e_{abc}(t)$ is the vector of line-to-neutral back-EMF voltages, found as
\begin{equation}
    e_{abc}(t) \triangleq \begin{bmatrix} e_a(t) \\ e_b(t) \\ e_c(t) 
    \end{bmatrix} = \begin{bmatrix} \sin(\theta_{re}(t)) \\ \sin\left(\theta_{re}(t) - \frac{2\pi}{3}\right)  \\ \sin\left(\theta_{re}(t) + \frac{2\pi}{3}\right)  \end{bmatrix} \Lambda_{PM} \dot{\theta}_{re}
\end{equation}
where $\Lambda_{PM}$ is the permanent-magnet flux linkage and $\theta_{re}(t) \triangleq N_p \theta_r(t) /2$ is the electrical rotor angle with $N_p$ being the number of poles of the machine. 
A graphical representation of the three-phase electrical model of the PMSM is provided in Figure \ref{inverter}.
For the purposes of analysis and control, it is beneficial to project the three-phrase variables onto a reference frame that rotates with $\theta_{re}(t)$. 
This is accomplished using the combined Clarke/Park transformation \cite{park1929two} defined as
\begin{equation}
    P\left(\theta_{re}\right) \triangleq
    \setlength{\arraycolsep}{1.5pt}
    \frac{2}{3}\begin{bmatrix} \cos\left(\theta_{re}\right) & \cos\left(\theta_{re}-\frac{2\pi}{3}\right) & \cos\left(\theta_{re}+\frac{2\pi}{3}\right) \vspace{5pt}\\
    -\sin\left(\theta_{re}\right) & -\sin\left(\theta_{re} - \frac{2\pi}{3}\right) & -\sin\left(\theta_{re} + \frac{2\pi}{3}\right) \vspace{5pt}\\ 
    \frac{1}{2} & \frac{1}{2} & \frac{1}{2}
    \end{bmatrix}
\end{equation}
with the corresponding inverse transformation
\begin{equation}
    P^{-1}\left(\theta_{re}\right) =
    \begin{bmatrix} \cos\left(\theta_{re}\right) & -\sin(\theta_{re})& 1  \vspace{5pt}\\
    \cos\left(\theta_{re} - \frac{2\pi}{3}\right) & -\sin\left(\theta_{re} - \frac{2\pi}{3}\right) & 1 \vspace{5pt}\\ 
     \cos\left(\theta_{re} + \frac{2\pi}{3}\right) & -\sin\left(\theta_{re} + \frac{2\pi}{3}\right) & 1
    \end{bmatrix}
\end{equation}
and where we have suppressed the time-dependency of $\theta_{re}(t)$.
Subsequently, we define vectors
\begin{align}
    i_{dq0}^r(t) \triangleq& \begin{bmatrix} i_d^r(t) \\ i_q^r(t) \\ i_0^r(t)
    \end{bmatrix} = P\left(\theta_{re}(t)\right)i_{abc}(t)\\
    v_{dq0}^r(t) \triangleq& \begin{bmatrix} v_d^r(t) \\ v_q^r(t) \\ v_0^r(t)
    \end{bmatrix} = P\left(\theta_{re}(t)\right)v_{abc}(t)
\end{align}
where the subscripts $dq0$ refer to the direct-axis, quadrature-axis, and zero components, respectively and the superscript $r$ denotes the rotor reference frame. 
It follows that the dynamics of $i_{dq0}^r(t)$ evolve according to
\begin{multline}
    \label{r_eqn_init}
    P(\theta_{re}(t))\frac{d}{dt}\bigg(P^{-1}(\theta_{re}(t))i_{dq0}^r(t) \bigg) \\
    = \frac{1}{L} \big( v_{dq0}^r(t) - R i_{dq0}^r(t) + P(\theta_{re}(t))e_{abc}(t) \big)
\end{multline}
Expanding \eqref{r_eqn_init} and making the substitution $\dot{\theta}_{re}(t) = \frac{N_p}{2\ell}\dot{x}(t)$, we obtain the following system of coupled differential equations
\begin{align}
    \frac{d}{dt}i_d^r(t) &= \frac{1}{L}\left(v_d^r(t) -Ri_d^r(t) +\frac{N_p}{2\ell}\dot{x}(t)L i_q^r(t)  \right) \label{id_eqn} \\
    \frac{d}{dt}i_q^r(t) &= \frac{1}{L}\left(v_q^r(t) -Ri_q^r(t) -\frac{N_p}{2\ell}\dot{x}(t)(L i_d^r(t) + \Lambda_{PM})  \right) \label{iq_eqn} \\
     \frac{d}{dt}i_0^r(t) &= \frac{1}{L}\left(v_0^r(t) -Ri_0^r(t)  \right) 
\end{align}
We assume that the three-phase windings are connected in an ungrounded wye configuration, implying that $i_0^r(t) = \frac{1}{3}\left(i_a(t)+i_b(t)+i_c(t)\right)  = 0~\forall t$, due to Kirchoff's current law applied to the neutral node. 
From this it follows that $v_0^r(t) = 0 ~\forall t$. 

Finally, it can be shown \cite{krishnan2017permanent} that the electromechanical force is proportional to the quadrature-axis current and is given by
\begin{equation} \label{fe_iq_relationship}
    f_e(t) = \frac{3N_p \Lambda_{PM}}{4\ell}i_q^r(t)
\end{equation}

\subsection{Power generation}
The instantaneous electrical power delivered to the transducer is defined as
\begin{align}
    P_{elec}(t) \triangleq & v_{abc}^T(t) i_{abc}(t) \\
    =& \tfrac{3}{2} v_{dq0}^{rT}(t) i_{dq0}^r(t) \\
    =& \tfrac{3}{2} \left( v_d^r(t) i_d^r(t) + v_q^r(t) i_q^r(t) \right)
 \end{align}
where positive $P_{elec}$ implies conversion of electrical to mechanical energy (motoring). Accordingly, the power generated by the energy harvester is defined as $P_{gen} \triangleq -P_{elec}$. 

\subsection{Combined electromechanical state space}

Assembling \eqref{sdof_eom}, \eqref{f_complete}, \eqref{id_eqn}, \eqref{iq_eqn}, \eqref{fe_iq_relationship} into state-space form provides a complete representation of the nonlinear  electromechanical dynamics of the energy harvester, as
\begin{equation}
    \frac{d}{dt} \begin{bmatrix}
    x \\ \dot{x} \\ i_d^r \\ i_q^r \end{bmatrix} =
    \begin{bmatrix}
         \dot{x} \\
         -\frac{k}{m}x-\frac{c}{m}\dot{x}-a+\frac{1}{m}\Phi\left( x,\dot{x},\frac{3N_p \Lambda_{PM}}{4\ell}i_q^r,a\right) \\
         -\frac{R}{L} i_d^r + \frac{N_p}{2\ell}\dot{x}i_q^r + \frac{1}{L} v_d^r \\
         -\frac{N_p\Lambda_{PM}}{2L\ell}\dot{x}-\frac{R}{L} i_q^r -\frac{N_p}{2\ell}\dot{x} i_d^r + \frac{1}{L} v_q^r
    \end{bmatrix}
\end{equation}
where we have suppressed the time-dependency of the state variables, and disturbance $a$.

\subsection{Effect of finite bus voltage on current feasibility}

We assume that the $i_{abc}$ currents (and consequently $i_q^r$ and $i_d^r$) are regulated at high-bandwidth (at least two decades beyond the energy harvester's natural frequency) via proportional-integral (PI) feedback control. 
As discussed, the power electronic drive in Figure~\ref{inverter} uses high-frequency PWM of the PMSM line-to-line voltages to realize the PI commands. 
Using a time-scale separation argument, it follows that $\{i_q^r,i_d^r\}$ may be viewed as control inputs from the perspective of the mechanical system dynamics. 
However, at a given time $t$, the feasibility of a desired $\{i_q^{r}(t),i_d^{r}(t)\}$ pair depends on the inverter's bus voltage $V_s$.
Assuming that a simple sinusoidal PWM scheme is used by each of the drive's three half-bridges, it can be shown (see e.g., \cite{mohan2012electric}) that the maximum magnitude of any three-phase line-to-neutral voltage is $V_s/2$. 
In balanced operation, we have that 
\begin{align}
    v_{an}(t) &= v_{ph}(t) \sin(\theta_{re}(t)+\phi) \\ 
    v_{bn}(t) &= v_{ph}(t) \sin(\theta_{re}(t)-\frac{2\pi}{3}+\phi)\\
    v_{cn}(t) &= v_{ph}(t)\sin(\theta_{re}(t)+\frac{2\pi}{3}+\phi)
\end{align}
where $v_{ph}(t)$ is the voltage amplitude and $\phi$ is an arbitrary constant phase angle. 
Consequently, this implies
\begin{align} \label{abc_ineq}
    \sqrt{v_{abc}^T(t) v_{abc}(t)}
    =& \sqrt{v_{an}^2(t) + v_{bn}^2(t) + v_{cn}^2(t) } \\
    =& \sqrt{\tfrac{3}{2}} |v_{ph}(t)| \\
    \leq & \sqrt{\tfrac{3}{2}} \frac{V_s}{2}
\end{align}
But
\begin{align} \label{dq0_ineq_1}
    \sqrt{v_{abc}^T(t) v_{abc}(t) } &= \sqrt{v_{dq0}^T(t) P^{-T}(t) P^{-1}(t) v_{dq0}(t) } \nonumber \\ &=\sqrt{\tfrac{3}{2}(v_{d}^{r2}(t) + v_{q}^{r2}(t)) } 
\end{align}
Combining \eqref{abc_ineq} and \eqref{dq0_ineq_1} we see that the rotor reference frame voltages must satisfy
\begin{equation} \label{dq0_ineq_2}
    \sqrt{v_{d}^{r2}(t) + v_{q}^{r2}(t)} \leq \frac{V_s}{2} ~~ \forall t
\end{equation}
We note that if the more complex space vector modulation (SVM) algorithm is used to perform PWM, the right-hand-side of \eqref{dq0_ineq_2} would increase to $V_s/\sqrt{3}$ (see e.g., \cite{krishnan2017permanent} for details). 

Next, suppose $\dot{x}(t),~v_q^r(t),$ and $v_d^r(t)$ are held constant. 
In steady-state we have $\frac{d}{dt}i_d^r(t) = \frac{d}{dt}i_q^r(t) = 0$ implying the static relationships
\begin{align} \label{v_d_eqn}
    v_d^r =& Ri_d^r -\frac{N_p}{2l}\dot{x}L i_q^r
\\
\label{v_q_eqn}
    v_q^r =& Ri_q^r +\frac{N_p}{2l}\dot{x}(L i_d^r + \Lambda_{PM})
\end{align}
Substituting these expressions back into the left-hand side of \eqref{dq0_ineq_2} and then simplifying, we obtain a feasibility condition for the currents $\{i_q^r,i_d^r\}$
\begin{multline} \label{idq_constraint}
    \left(R^2 + \frac{N_p^2 L^2}{4\ell^2}\dot{x}^2\right)\left(i_q^{r2} + i_d^{r2} \right)   +  \frac{N_p\Lambda_{PM}}{\ell}\dot{x} \\ \times \left(R i_q^r  + \frac{N_pL}{2\ell}\dot{x}i_d^r\right) 
     \leq \frac{1}{4} \left(V_s^2 - \frac{N_p^2\Lambda_{PM}^2}{\ell^2}\dot{x}^2\right)
\end{multline}
It is important to note that this constraint is quasi-static. Following the approach taken in \cite{scruggs2018analysis}, we  approximately account for the dynamic behavior of $\dot{x}(t),~i_q^r(t),$ and $i_d^r(t)$, by tightening \eqref{idq_constraint} using the safety factor $\delta < 1$ as follows
\begin{multline} \label{idq_delta_constraint}
    \left(R^2 + \frac{N_p^2 L^2}{4\ell^2}\dot{x}^2\right)\left(i_q^{r2}(t) + i_d^{r2}(t) \right)  \\
    + \frac{N_p\Lambda_{PM}}{\ell}\dot{x}(t) \left(R i_q^r(t)  + \frac{N_pL}{2\ell}\dot{x}(t)i_d^r(t)\right) \\
     \leq \frac{1}{4} \left((\delta V_s)^2 - \frac{N_p^2\Lambda_{PM}^2}{\ell^2}\dot{x}^2(t)\right)
\end{multline}
Next, observe that \eqref{idq_delta_constraint} is equivalent to 
\begin{multline}\label{idq_delta_constraint_2}
    \left(i_q^r(t) + \frac{2 N_p \Lambda_{PM}R \ell \dot{x}(t)}{(2R\ell)^2 + (N_p L \dot{x}(t))^2} \right)^2  \\ + \left(i_d^r(t) + \frac{(N_p\dot{x})^2(t) \Lambda_{PM}L}{(2R\ell)^2 + (N_p L \dot{x}(t))^2}  \right)^2  \\
    \leq \frac{(\delta \ell V_s)^2}{(2R\ell)^2 + (N_p L \dot{x}(t))^2}
\end{multline}
In this form, it becomes clear that in order for there to exist an $i_d^r(t)$ that satisfies the constraint, it is necessary that $i_q^r(t)$ first satisfy
\begin{equation}
    i_q^r(t) \in \left[ I_q^{\min}(\dot{x}(t)) , I_q^{\max}(\dot{x}(t)) \right]
\end{equation}
where
\begin{align} \label{i_q_1}
    I_q^{\max}(\dot{x}) =& \frac{\delta \ell V_s}{\sqrt{(2R\ell)^2 + (N_p L \dot{x})^2}} - \frac{2 N_p \Lambda_{PM}R \ell \dot{x}}{(2R\ell)^2 + (N_p L \dot{x})^2}
\\
\label{i_q_2}
    I_q^{\min}(\dot{x}) =& -\frac{\delta \ell V_s}{\sqrt{(2R\ell)^2 + (N_p L \dot{x})^2}} - \frac{2 N_p \Lambda_{PM}R \ell \dot{x}}{(2R\ell)^2 + (N_p L \dot{x})^2}
\end{align}

\subsection{Current rating constraints}

In addition to current constraint \eqref{idq_delta_constraint_2} arising from the finite bus voltage, PMSMs (and electric machines in general) will have continuous and peak current ratings denoted $i_{cont}$ and $i_{peak}$, respectively. In general, the former may be exceeded briefly during
operation, while the latter should not be exceeded to avoid damaging the device and creating a safety hazard. Obviously, both ratings could be satisfied by imposing
\begin{equation} \label{idq_rating}
    \sqrt{i_d^{r2} + i_q^{r2}} \leq i_{cont} < i_{peak} ~~ \forall t
\end{equation}
However, this would be overly conservative, and we will discuss how it might be appropriately relaxed later in the paper.

\subsection{Disturbance model} 
We assume the stochastic disturbance $a(t)$ has a second-order bandpass spectrum and is modeled the output of a filter with state space representation
\begin{equation}
\begin{array}{rl}
\frac{d}{dt}\begin{bmatrix} d(t) \\ a(t) \end{bmatrix} =& \begin{bmatrix} 0 & 1 \\ -\omega_a^2 & -2\zeta_a \omega_a \end{bmatrix} \begin{bmatrix} d(t) \\ a(t) \end{bmatrix} + \begin{bmatrix} 0 \\ 2\sigma_a \sqrt{\zeta_a \omega_a} \end{bmatrix} w(t)
\end{array} 
\end{equation}
where $w(t)$ is a scalar, stationary, white noise process with zero mean and unit spectral intensity (i.e., $\Ex\{w(t)w(\tau)\}=\delta(t-\tau)$), $d(t)$ is an internal dynamic state, $\omega_a$ is the passband frequency, $\sigma_a$ is the disturbance intensity, and $\zeta_a$ is the damping ratio.

\subsection{Augmented state space}
Proceeding with the assumption that $\{i_q^r(t),i_d^r(t)\}$ may be considered as control inputs, we combine the energy harvester dynamics and the disturbance model to obtain the following augmented system $\mathcal{S}$, with state space representation
\begin{equation} \label{nonlin_state_space}
    \mathcal{S} : \left\{ \begin{array}{rl}
        \tfrac{d}{dt} \xi(t) &= \Psi\left( \xi(t), i_q^r(t) \right) + B_w w(t) \\
        y(t) &= C_y \xi(t) + n(t) 
    \end{array} \right.
\end{equation}
where state vector $\xi \triangleq \begin{bmatrix} x & \dot{x} & d & a \end{bmatrix}^T$, vector $y$ contains the measured outputs, and 
\begin{align}
    \Psi\left( \xi, i_q^r \right) &\triangleq \begin{bmatrix}
        \dot{x} \\
        -\frac{k}{m}x-\frac{c}{m}\dot{x}-a+\frac{1}{m}\Phi\left( x,\dot{x},\frac{3N_p \Lambda_{PM}}{4\ell}i_q^r,a\right) \\
        a \\
        -\omega_a^2 d - 2\zeta_a \omega_a a
    \end{bmatrix}
    \\
    B_w &\triangleq \begin{bmatrix} 0 &
    0 &
    0 & 2\sigma_a \sqrt{\zeta_a \omega_a} \end{bmatrix}^T
\end{align}
where we have suppressed the time-dependency of $\xi(t)$ and $i_q^r(t)$.
We assume that $y(t)$ is corrupted by a white noise vector $n(t)$, which has zero mean and intensity $\Phi_n$. We further assume that $y(t)$ contains a noise-corrupted version of the transducer velocity, which we denote $\dot{\tilde{x}}(t)$; i.e., that there exists a matrix $T_{vy}$ such that $\dot{\tilde{x}}(t) = T_{vy}y(t)$.

\section{Control synthesis}

In this section, we present a heuristic method for designing a dynamic output-feedback control law $\mathcal{K}:y\mapsto\{i_q^r,i_d^r\}$ that approximately maximizes the average power generated by the energy harvester in stationarity and ensures the feasibility constraint \eqref{idq_delta_constraint_2} is satisfied at all times. The proposed controller $\mathcal{K}$ will actually consist of two distinct, but coupled, feedback laws $\mathcal{K}_q:y \mapsto i_q^r$ and $\mathcal{K}_d:\{y,\,i_q^r\} \mapsto i_d^r$. We will first design the quadrature-axis current controller $\mathcal{K}_q$ to maximize power generation via a multi-objective convex optimization, and then utilize the direct-axis current controller $\mathcal{K}_d$ to enforce \eqref{idq_delta_constraint_2} through field-weakening.

\subsection{Performance objective}

Given the stochastic nature of the disturbance, we seek to maximize the mean power generated by the harvester, defined as $\bar{P}_{gen}\triangleq\Ex\{P_{gen}\}$, where $\Ex\{\cdot\}$ denotes expectation in stationarity. Expanding this we have
\begin{align} \label{p_gen_eqn}
    \bar{P}_{gen} =& -\Ex \left\{ \tfrac{3}{2}  \left( v_d^r i_d^r + v_q^r i_q^r \right) \right\}  \\
    =&-\tfrac{3}{2} \left(  R\Ex \left\{i_d^{r2}+i_q^{r2}\right\} +C\Ex \left\{\xi i_q^r \right\} \right)
\end{align}
where matrix $C \triangleq \begin{bmatrix}
0 &\frac{N_p\Lambda_{PM}}{2\ell} & 0 & 0\end{bmatrix}$.

\subsection{General optimization formulation}
With the performance objective defined, we can now state the control synthesis problem in terms of the following nonconvex optimization problem

\begin{equation*} \label{nonconvex_opt}
\textrm{OP1}: \left\{ \begin{array}{rl}\begin{array}{lll}
\text{Given} &:& \mathcal{S}, R, V_s, \delta \\
\text{Maximize} &:& \bar{P}_{gen} \\
\text{Over} &:& \mathcal{K} \\
\text{Subject to} &:& \eqref{idq_delta_constraint_2}, \eqref{idq_rating}
\end{array}
\end{array} \right.
\end{equation*}
Solving this problem exactly is extremely challenging and remains an open research question. Here, we will only solve it approximately.

\subsection{Linear case with infinite $V_s$}
First, consider the simplified case in which the transducer's linear-to-rotational conversion mechanism is perfectly efficient, there is no Coulomb friction, and the drive bus voltage is infinitely large (i.e., $\eta=1,~f_c=0,$ and $V_s = \infty$). In this scenario, $\mathcal{S}$ becomes a linear system, i.e.,
\begin{equation}
    \mathcal{S} : \left\{ \begin{array}{rl}
        \tfrac{d}{dt} \xi(t) =& A\xi(t) + B i_q^r(t) + B_w w(t) \\
        y(t) =& C_y \xi(t) + n(t) 
    \end{array} \right.
\end{equation}
where
\begin{align}
\label{A_def}
    A =& \begin{bmatrix} 
        0 & 1 & 0 & 0 \\
        -\frac{k}{\tilde{m}} & -\frac{\tilde{c}}{\tilde{m}} & 0 & -\frac{m}{\tilde{m}} \\
        0 & 0 & 0 & 1 \\
        0 & 0 & -\omega_a^2 & -2\omega_a\zeta_a 
    \end{bmatrix}  \\
    \label{B_def}
    B =& \begin{bmatrix} 0 & \frac{3N_p\Lambda_{PM}}{4\ell \tilde{m}} & 0 & 0 \end{bmatrix}^T 
\end{align}
and where
\begin{align}
    \tilde{m} =& m + \frac{J}{\ell^2} \\
    \tilde{c} =& c + \frac{B}{\ell^2} 
\end{align}
Because $V_s = \infty$, \eqref{idq_delta_constraint_2} is automatically satisfied for all $\{i_q^r,i_d^r\}$ currents. 
As $i_d^r$ has no effect on the energy harvester's mechanical dynamics, it is optimal to control $i_d^r=0~\forall t$, in order to minimize resistive power losses. 
Consequently, \eqref{idq_rating} is reduced to
\begin{equation} \label{iq_rating}
    |i_q^r(t)| \leq i_{cont} ~~ \forall t
\end{equation}

In practice it is physically possible to exceed $i_{cont}$ for brief periods without adverse effects.
As such, we replace \eqref{iq_rating} with a constraint on the variance of $i_q^r$ , i.e.,
\begin{equation} \label{iq_var_cont}
    \Ex\{i_q^{r2}\} \leq \frac{1}{4} i_{cont}^2
\end{equation}
We justify the use of \eqref{iq_var_cont} to approximately constrain the peak values of $i_q^r$ as follows. 
Assume $\mathcal{K}_q$ is linear and let $\psi$ be a randomly-selected peak of the closed-loop, stationary response of $i_q^r$.
If $i_q^r$ is a narrowband process, then $\psi$ is Rayleigh-distributed (for details see e.g., \cite{cartwright1956statistical}). 
Then we have
\begin{align}
    \Pr(\psi \leq i_{cont} ) &= 1 - \exp \left( \frac{-i_{cont}^2}{2\sigma_i^2} \right) \\
    &\geq 1 - \exp \left( \frac{-i_{cont}^2}{2( i_{cont}^2/4)} \right) \\
    &= 1-\exp (-2) \approx 0.14
\end{align}
where $\sigma_i^2 \triangleq  \Ex\{i_q^{r2}\}$. 
It follows that any linear controller adhering to \eqref{iq_var_cont} ensures that the majority (i.e., $\geq $86\%) of the current peaks are below the continuous current rating.

To further enhance the tractability of the control design problem, we restrict the optimization domain of $\mathcal{K}_q$ to LTI, strictly proper transfer functions, with state space realizations of dimension equal to that of $\mathcal{S}$. 
As such, we presume
\begin{equation} \label{q_controller}
    \mathcal{K}_q: \left\{ \begin{array}{rl} \frac{d}{dt}x_K(t) &= A_K x_K(t) + B_K y(t) \\
    i_q^r &= C_K x_K(t)
    \end{array} \right.
\end{equation}
where $\textrm{dim}(x_K)=\textrm{dim}(\xi)$, and seek to optimize the triple $\{A_K,B_K,C_K\}$. 

With these assumptions made, OP1 can be rewritten as a convex, semi-definite program by applying the following theorem.

\begin{th1}\label{theorem_1}
Let $\eta=1,~f_c=0,V_s = \infty$, and $i_d^r(t)=0~\forall t$. 
There exists a stabilizing LTI feedback law $\mathcal{K}_q:y\mapsto i_q^r$ as in \eqref{q_controller}, such that $\bar{P}_{gen} > \gamma$ and $\sigma_i^2< \frac{1}{4} i_{cont}^2$, if and only if there exist dimensionally-compatible matrices $X=X^T$, $Y=Y^T$, $\tilde{A}$, $\tilde{B}$, $\tilde{C}$, and scalar $\beta$ such that
\begin{align}
    \begin{bmatrix} \Delta_1+\Delta_1^T & A+\tilde{A}^T & B_w & 0 \\ \star & \Delta_2+\Delta_2^T & YB_w & \tilde{B} \\ \star & \star& -I & 0 \\ \star & \star & \star & -\Phi_n^{-1}  \end{bmatrix} <& 0 \label{lmi1}\\
    \begin{bmatrix}  \frac{1}{4}i_{cont}^2 & \tilde{C} & 0 \\ \star & X & I \\ \star & \star & Y\end{bmatrix} >& 0 \label{lmi2}\\
    \begin{bmatrix} 
    \beta & \tilde{C}-HX& -H \\ \star & X& I \\ \star & \star & Y
    \end{bmatrix} >& 0 \label{lmi3} \\
    -\tfrac{3}{2}\left(\tfrac{1}{2}B_w^T S B_w + \beta R \right) >& \gamma \label{lmi4} 
\end{align}
where 
\begin{equation}
\Delta_1 \triangleq AX+B\tilde{C}, \quad \Delta_2 \triangleq YA+\tilde{B}C_y
\end{equation}
and where $H=-\frac{1}{2}R^{-1} (B^T S + C)$, and $S = S^T$ is the unique stabilizing solution to Riccati equation
\begin{equation}
    A^T S + S A - \frac{1}{2}(SB+C^T) R^{-1} (B^T S+C) = 0.
\end{equation}
Furthermore, if the above inequalities are feasible, then one such controller is obtained via
\begin{align}
A_K =& N^{-1} \left[ \tilde{A} -YAX - \tilde{B} C_y X - YB\tilde{C} \right] M^{-T} \label{A_eqn} \\
B_K =& N^{-1} \tilde{B} \label{B_eqn} \\
C_K =& \tilde{C} M^{-T} \label{C_eqn}
\end{align}
and $M$ and $N$ are any matrices that satisfy $XY +MN^T = I$.
\end{th1}
\begin{proof}
See \cite{Scruggs2012} for an analogous proof. It uses standard linear matrix inequality (LMI) techniques described by \cite{Scherer1997}.
\end{proof}

We obtain the optimal $\mathcal{K}_q$ by solving the convex optimization
\begin{equation*} \label{convex_opt_1}
\textrm{OP2}: \left\{ \begin{array}{rl}\begin{array}{lll}
\text{Given} &:& \mathcal{S}, R\\
\text{Minimize} &:& -\gamma \\
\text{Over} &:& \gamma,\beta, \tilde{A},\tilde{B},\tilde{C}, \\ & &  X=X^T, Y=Y^T\\
\text{Subject to} &:& \eqref{lmi1}, \eqref{lmi2}, \eqref{lmi3}, \eqref{lmi4}
\end{array}
\end{array} \right.
\end{equation*}
and then computing $\{A_K,B_K,C_K\}$ via the inversion of equations \eqref{A_eqn}, \eqref{B_eqn}, and \eqref{C_eqn} .

\subsection{Accounting for finite $V_s$ in the design of $\mathcal{K}_q$} \label{finite_section}

Next, we consider the case in which $V_s$ is finite and constraint \eqref{idq_delta_constraint_2} must be satisfied. 
First, observe that as $|\dot{x}(t)|$ becomes larger, the set of feasible $\{i_q^r(t),i_d^r(t)\}$ currents shrinks. 
It follows that one potential way to reduce the possibility of $\mathcal{K}_q$ producing infeasible $i_q^r$ currents is to impose a constraint on the closed-loop, mean-square response of the transducer velocity, i.e.,
\begin{equation} \label{velocity_con}
    \Ex\{\dot{x}^2\} < \dot{x}_{m}^2
\end{equation} 
where $\dot{x}_{m} > 0$ is some constant. Using the variables introduced in Theorem \ref{theorem_1}, it can be shown (see e.g., \cite{robert2017unified}) that \eqref{velocity_con} is equivalent to the following LMI
\begin{equation}
     \begin{bmatrix}  \dot{x}_{m}^2 & C_v X & C_v \\ \star & X & I \\ \star & \star & Y\end{bmatrix} > 0 \label{lmi5}
\end{equation}
where
\begin{align}
    C_v \triangleq & \begin{bmatrix} 0 & 1 & 0 & 0 \end{bmatrix} 
\end{align}
Obviously, any $\mathcal{K}_q$ adhering to \eqref{lmi5} could still produce infeasible $i_q^r(t)$ commands that violate \eqref{i_q_1} and \eqref{i_q_2}, compromising the closed-loop system behavior. 
However, we can significantly reduce the probability of this happening by introducing another constraint into the optimization of $\mathcal{K}_q$. 

We begin by noting that \eqref{i_q_1} and \eqref{i_q_2} are equivalent to
\begin{multline} \label{nonconvex_i_q}
    i_q^{r2}(t)\left((2R\ell)^2+(N_pL\dot{x}(t))^2 \right) +4N_p\Lambda_{PM}R\ell\dot{x}(t)i_q^r(t) \\ + \frac{(2N_p\Lambda_{PM}R\ell\dot{x}(t))^2}{(2R\ell)^2+(N_pL\dot{x}(t))^2} 
    \leq (\delta \ell V_s)^2
\end{multline}
This constraint is unfortunately nonconvex. 
However, \eqref{nonconvex_i_q} can be conservatively satisfied (at most points in time) by imposing 
\begin{multline}
    i_q^{r2}(t)\left(4R^2\ell^2+N_p^2L^2\dot{x}_m^2 \right) +4N_p\Lambda_{PM}R\ell\dot{x}(t)i_q^r(t) \\ + \Lambda_{PM}^2 N_p^2 L^2 \dot{x}^2(t) \leq (\delta \ell V_s)^2
\end{multline} 
which in turn is equivalent to
\begin{equation} \label{i_q_conservative}
    \left(i_q^r(t) R +\dot{x}(t)\frac{\Lambda_{PM}N_p}{2\ell}\right)^2 + \left(i_q^{r}(t)\dot{x}_m\frac{N_p L}{2\ell}\right)^2 \leq \left(\frac{\delta V_s}{2}\right)^2
\end{equation}
Using the same reasoning as in Section III-C, we enforce the above constraint in a relaxed probabilistic sense, i.e.,
\begin{equation} \label{i_q_stochastic}
    \Ex \left\{\left(i_q^r R +\dot{x}\frac{\Lambda_{PM}N_p}{2l}\right)^2 + \left(i_q^{r}\dot{x}_m\frac{N_p L}{2l}\right)^2 \right\} < \frac{1}{4}\left(\frac{\delta V_s}{2}\right)^2
\end{equation}
which is convex and again can be rewritten as an LMI
\begin{multline}
\setlength{\arraycolsep}{3pt}
    \begin{bmatrix} \frac{1}{4}(\frac{\delta V_s}{2})^2  & R \tilde{C}+\frac{\Lambda_{PM} N_p}{2 \ell} C_v X  &   \frac{\Lambda_{PM} N_p}{2 \ell} C_v  &  \frac{\dot{x}_m N_p L}{2 \ell} \tilde{C} &  0 \\
         \star & X & I & 0 & 0\\
         \star & \star & Y & 0 & 0 \\
         \star & \star & \star & X & I \\
         \star & \star & \star & \star &  Y  \end{bmatrix} \\ > 0 \label{lmi6}
\end{multline}
Incorporating \eqref{lmi5} and \eqref{lmi6} into the $\mathcal{K}_q$ optimization problem, we obtain
\begin{equation*} \label{convex_opt_2}
\textrm{OP3}: \left\{ \begin{array}{rl}\begin{array}{lll}
\text{Given} &:& \mathcal{S}, R\\
\text{Minimize} &:& -\gamma \\
\text{Over} &:& \gamma,\beta, \tilde{A},\tilde{B},\tilde{C}, \\ & &  X=X^T, Y=Y^T\\
\text{Subject to} &:& \eqref{lmi1}, \eqref{lmi2}, \eqref{lmi3}, \eqref{lmi4}, \eqref{lmi5}, \eqref{lmi6}
\end{array}
\end{array} \right.
\end{equation*}
which maintains the convexity of OP2. By designing $\mathcal{K}_q$ to adhere to both \eqref{lmi5} and \eqref{lmi6}, we ensure that approximately $\geq$86\% of the $i_q^r$ current peaks satisfy \eqref{i_q_conservative}. 
Obviously, this is not sufficient to guarantee $i_q^r(t)$ feasibility at all times, and we therefore impose a secondary ``clipping" action using the upper and lower bounds on $i_q^r$ provided by \eqref{i_q_1} and \eqref{i_q_2}. 
Letting $i_q^{r*}(t)$ denote the control input produced by $\mathcal{K}_q$, we implement the dynamic saturation
\begin{equation} \label{sat_eqn}
    i_q^r(t) = \underset{{[I_q^{\min}(\dot{x}(t)), I_q^{\max}(\dot{x}(t))]}}{\textrm{sat}}\{i_q^{r*}(t)\}
\end{equation}
where $I_q^{\max}(\cdot)$ and $I_q^{\min}(\cdot)$ are defined by \eqref{i_q_1} and \eqref{i_q_2}, respectively.

We also note that OP3 does not consider the impact of the $i_d^r$ current on the performance objective $\bar{P}_{gen}$. In fact, the parameter $\gamma$ serves only as an \emph{approximate} lower bound on the quantity $\left(\bar{P}_{gen}+\frac{3}{2}R\Ex\{i_d^{r2}\}\right)$. We say it is an approximate bound because it does not account for the saturation introduced by \eqref{sat_eqn}.

\begin{table}
\caption{Oscillator and Disturbance Characteristics}
\centering
\small
\label{table:1}
\begin{tabular}{l c} 
Parameter & Value\\ 
\hline\hline
SDOF mass ($m$) & 3000 kg \\ 
SDOF stiffness ($k$) & 1.1844$\times 10^5$ N-m$^{-1}$ \\
SDOF viscous damping ($c$) & 942.47 N-s-m$^{-1}$ \\
Disturbance passband frequency ($\omega_a$) & $2\pi$ rad-s$^{-1}$ \\
Disturbance damping ratio ($\zeta_a$) & 0.1 \\
\hline
\end{tabular}
\end{table}

\begin{table}
\caption{Transducer Characteristics}
\centering
\small
\label{table:2}
\begin{tabular}{l c} 
Parameter & Value\\ 
\hline\hline
Resistance ($R$) & 10.7 $\Omega$ \\ 
Inductance ($L$) & 0.0219 H \\
Permanent-magnet flux linkage ($\Lambda_{PM}$) & 0.1603 V-s \\
No. of poles ($N_p$) & 6 \\
Rotational inertia ($J$) & 3.54 $\times 10^{-5}$ kg-m$^2$\\ 
Rotational viscous damping ($B$) & 3.25 $\times 10^{-4}$ N-m-s\\ 
Coulomb friction ($f_c$) & 35 N\\ 
Lead length ($l$) & 2.55 $\times 10^{-3}$ m-rad$^{-1}$\\
Efficiency ($\eta$) & 0.91\\ 
Continuous current rating ($i_{cont}$) & 2 A \\
\hline
\end{tabular}
\end{table}

\subsection{Design of $\mathcal{K}_d$}

We now turn our attention to the design of $\mathcal{K}_d$. 
The direct-axis current $i_d^r(t)$ has no effect on the mechanical dynamics of the energy harvester. 
In addition, nonzero $i_d^r(t)$ reduces power generation by increasing resistive power losses. 
It is therefore optimal to control $i_d^r(t)=0$, unless this leads to violation of \eqref{idq_delta_constraint_2}. 
In this case, it is desirable to make the magnitude of $i_d^r(t)$ as small as possible while still satisfying the constraint, in order to minimize losses. Accordingly, we implement the feedback law $\mathcal{K}_d:\{y,\,i_q^r\}  \mapsto i_d^r$ as
\begin{equation} \label{d_controller}
   i_d^r(t) = \min \left\{0, \sigma(\dot{\hat{x}}(t),i_q^r(t))-\frac{(N_p\dot{\hat{x}}(t))^2 \Lambda_{PM}L}{(2R\ell)^2 + (N_p L \dot{\hat{x}}(t))^2}\right \}  
\end{equation}
where $\dot{\hat{x}}(t)$ is a low-pass filtered version of the noise-corrupted, measured transducer velocity $\dot{\tilde{x}}(t)$ and

\begin{multline}
    \sigma(\dot{\hat{x}},i_q^r) \triangleq \\ \sqrt{\frac{(\delta \ell V_s)^2}{(2R\ell)^2 + (N_p L \dot{\hat{x}})^2} -  \left(i_q^r + \frac{2 N_p \Lambda_{PM}R \ell \dot{\hat{x}}}{(2R\ell)^2 + (N_p L \dot{\hat{x}})^2} \right)^2}
\end{multline}
and where we have suppressed the time-dependence of $\dot{\hat{x}}(t)$ and $i_q^r(t)$. Low-pass filtering of $\dot{\tilde{x}}(t)$ is performed to prevent the introduction and amplification of high-frequency noise in the $i_d^r$ commands. This filter should have a cutoff frequency well beyond the harvester and disturbance dynamics to avoid signal distortion in the frequency band of interest.

In effect, feedback law \eqref{d_controller} is used to counteract the back-EMF term $\frac{N_p \Lambda_{PM}}{2l} \dot{x}(t)$ in differential equation \eqref{iq_eqn} for $i_q^r(t)$. 
This strategy is known as \emph{field-weakening} in the literature, since a negative $i_d^r(t)$ current ``weakens" the magnetic field produced by the rotor magnets. 

\subsection{Linearization of transducer dynamics}
Finally, we remove the assumptions that $\eta=1$ and $f_c=0$. 
To make the $\mathcal{K}_q$ controller design analytically tractable, we first note that for the system considered here, it was found that the optimal feedback law resulted in the transducer being backdriven for the vast majority of the dynamic response, resulting in $p(t)<0$ for most $t$.
As such, for the purposes of control design, the function $h(p(t))$ can be approximated by $1/\eta$ for all $t$.
Doing so allows us to simplify $\Psi(\cdot)$ in \eqref{nonlin_state_space} by
\begin{equation}
    \Psi(\xi(t),i_q^r(t))
    = A \xi(t) + B i_q^r(t) + F \textrm{sgn}(\dot{x}(t))
\end{equation}
where $A$ and $B$ are as in \eqref{A_def} and \eqref{B_def} respectively, but with 
\begin{align}
    \tilde{m} =& m + \frac{J}{\eta \ell^2} \\
    \tilde{c} =& c + \frac{B}{\eta \ell^2} 
\end{align}
and $F$ is 
\begin{equation}
    F = \begin{bmatrix} 
        0 & -\frac{f_c}{\tilde{m}} & 0 & 0 
    \end{bmatrix}^T
\end{equation}
The validity of this assumption was verified in both simulation and experiment.

Next, we use stochastic linearization \cite{roberts2003random}
to address the Coulomb friction force. 
We assume that, in closed-loop, the augmented state
\begin{equation}
    \nu = \begin{bmatrix} \xi^T & x_K^T(t) \end{bmatrix}^T
\end{equation}
has a probability distribution $\phi(\nu)$ that can be approximated as Gaussian with zero mean and stationary covariance matrix $\Sigma = \Ex \{\nu \nu^T \}$, i.e., 
\begin{equation}
    \phi(\nu) \approx \frac{1}{\sqrt{(2\pi)^n \det \Sigma}}\exp\left\{-\frac{1}{2}\nu^T \Sigma \nu\right\}
\end{equation}
and then find the value of $\Sigma$ that brings about the weak stationarity condition
\begin{equation}
    \tfrac{d}{dt} \Ex\left\{ \nu(t)\nu^T(t) \right\} = 0
\end{equation}
This technique is also referred to as Gaussian closure, statistical linearization, equivalent linearization, and quasilinearization in the literature. 
It can be shown (see \cite{cassidy2012statistically} for details) that this results in the solution to the nonlinear, Lyapunov-like equation
\begin{equation} \label{lyap_like}
    A_{cl}(\Sigma)\Sigma + \Sigma A_{cl}^T(\Sigma) + B_{wcl} B_{wcl}^T = 0
\end{equation}
where 
\begin{equation} \label{A_cl_eqn}
    A_{cl}(\Sigma) \triangleq \begin{bmatrix} A_{eq}(\Sigma) & B C_K \\ B_K C_y & A_K \end{bmatrix}, ~~ 
        B_{wcl} = \begin{bmatrix} B_w \\ 0 \end{bmatrix}
\end{equation}
and where
\begin{equation} \label{V_eq}
    A_{eq}(\Sigma) = A + \sqrt{\frac{2}{\pi}}\frac{FC_v}{\sqrt{C_v\Sigma C_v^T}}
\end{equation}
We then have the stochastically-linearized plant model 
\begin{equation}
\mathcal{S}_{eq}: \left\{ \begin{array}{rl}
\frac{d}{dt}\xi(t) &= A_{eq}(\Sigma) \xi(t) + Bi_q^r(t) + B_w w(t) \\
    y(t) &= C_y \xi(t) + n(t)
\end{array} \right.
\label{lin_state_space}
\end{equation}
It is imperative to recognize that $\mathcal{S}_{eq}$ implicitly depends on $\mathcal{K}_q$ given the relationship between $A_{eq}$ and $\Sigma$. 
This dependence suggests that an iterative technique will be required to optimize $\mathcal{K}_q$.

\subsection{Iterative multi-objective optimization of $\mathcal{K}_q$} \label{iterative}

We propose the following procedure to optimize $\mathcal{K}_q$ for the stochastically-linearized system model:
\vspace{5pt}

\begin{itemize}
  \item[] \textbf{Step 0.} Set $A_{eq}=A$, and solve OP3 to obtain $\{A_K,B_K,C_K\}$.
  \item[] \textbf{Step 1.} Assemble $A_{cl}$ as in \eqref{A_cl_eqn} and compute $\Sigma$ by solving Lyapunov equation \eqref{lyap_like}.
  \item[] \textbf{Step 2.} Compute $A_{eq}(\Sigma)$ via \eqref{V_eq}.
  \item[] \textbf{Step 3.} Re-solve OP3 for the updated $\mathcal{S}_{eq}$ to obtain $\{A_K,B_K,C_K\}$ and $\gamma$. Return to Step 1.
\end{itemize}

\vspace{5pt}
Steps 1-3 are repeated until some convergence criterion on $\gamma$ is satisfied. Specifically, we use the absolute value of the change in $\gamma$ to assess convergence and cease iterating when $|\Delta \gamma| < 10^{-5}$. Although we offer no proof that this procedure is guaranteed to converge, we found that, for the examples studied in this paper, it generally converged within 20 iterations.

\subsection{Implementation of vector control scheme}

\begin{figure}
    \vspace{0pt}
    \centering
   \includegraphics[scale=0.58,trim={0cm 0cm 0cm 0cm},clip]{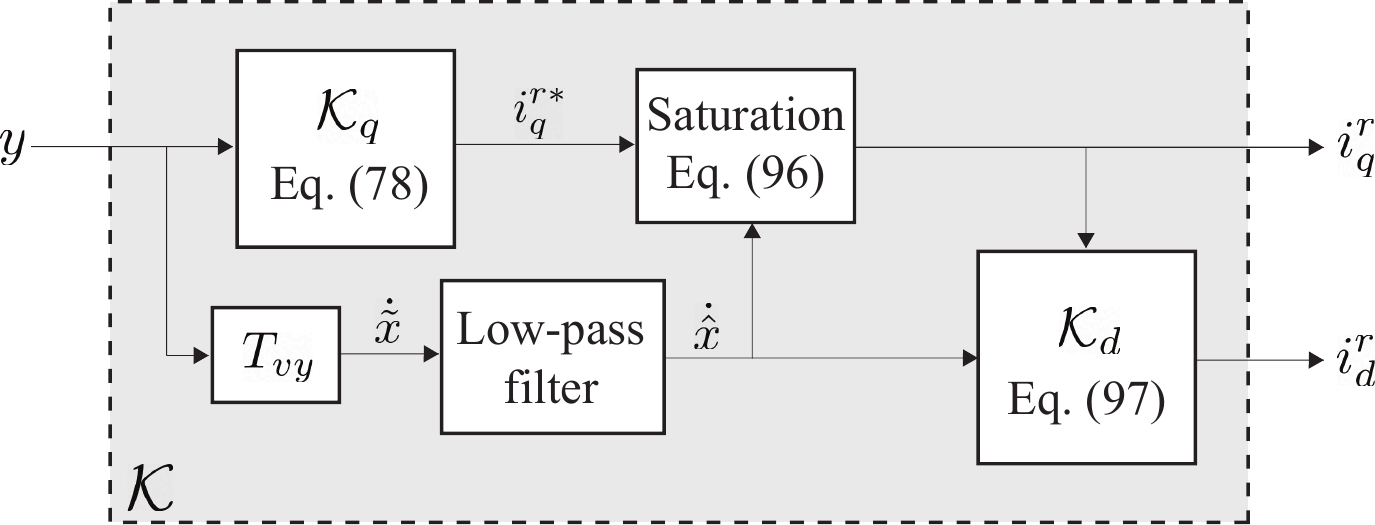}
   \caption{Block diagram of vector control law $\mathcal{K}$}
   \label{fig_K_block}
   \vspace{-10pt}
\end{figure}

Figure \ref{fig_K_block} provides an graphical representation of our vector control law $\mathcal{K}$. In summary, the quadrature-axis current controller $\mathcal{K}_q$, designed using the procedure proposed in the previous subsection, takes in feedback measurements $y$ and produces desired $i_q^{r*}$ commands. Then $i_q^{r*}$ is dynamically saturated according to \eqref{sat_eqn} using the filtered velocity measurement $\dot{\hat{x}}$ to produce feasible $i_q^r$. Subsequently, the direct-axis controller $\mathcal{K}_d$ given in \eqref{d_controller} produces the $i_d^r$ current needed to satisfy \eqref{idq_delta_constraint_2}. Next, the corresponding three-phase currents $i_{abc}$ are computed by applying the inverse Clarke/Park transform $P^{-1}(\theta_{re})$ to $i_{dq0}^r$ recalling that $i_0^r=0~\forall t$. Finally, a power electronic drive facilitates high-bandwidth tracking of the $i_{abc}$ commands, as described previously.

\section{Simulation}

Unfortunately, it is not possible to analytically compute the mean generated power $\bar{P}_{gen}$ associated with a $\mathcal{K}$ designed via our proposed methodology.
This is because the transducer mechanical dynamics are, in reality, nonlinear and our synthesis procedure exploits an approximate, linearized model. 
In addition, \eqref{sat_eqn} and $\eqref{d_controller}$ introduce additional nonlinearity, making it intractable to compute the expectations $\Ex\{i_q^{r2}+i_d^{r2}\}$ and $\left\{\xi i_q^r \right\}$ in \eqref{p_gen_eqn} analytically. 
Instead, it is necessary to assess the performance of a given $\mathcal{K}$ via numerical simulation, which accounts for all nonlinear effects.

In this section, we provide power generation results obtained via simulation for an energy harvester with the parameters listed in Tables \ref{table:1} and \ref{table:2}. We note that transducer characteristics given in Table \ref{table:2} correspond to an actual physical device, which is described in detail in Section \ref{exp_testbed_section}. We assume that only the transducer velocity is available for feedback (i.e., $C_y=C_v$). In addition, we presume the bus voltage $V_s=20$ V and set the safety factor $\delta=0.95$.

We specifically examined the effect of the transducer velocity constraint $\dot{x}_m$ and the disturbance intensity $\sigma_a$ on $\bar{P}_{gen}$. The convex programming software \emph{CVX} \cite{grant2008cvx} was used to perform the iterative controller optimization described in Section \ref{iterative} for each $\{\dot{x}_m,\sigma_a\}$ pair. The simulations were implemented in MATLAB/Simulink. We simulated the dynamic response of the full nonlinear system $\mathcal{S}$ over a time duration of 20 minutes, assuming instantaneous tracking of the current commands produced by $\mathcal{K}$ (i.e., we did not simulate the PWM switching of the power electronic drive nor the dynamics of the low-level PI current tracking loops). This assumption was validated by comparing to experimental results, as will shown subsequently. A running average was used to estimate the mean generated power, i.e.,
\begin{equation}
\\ \hat{\bar{P}}_{gen}(t)= -\frac{1}{t}\int_{0}^{t} \frac{3}{2} \left( v_d^r (\tau) i_d^r (\tau) + v_q^r (\tau) i_q^r (\tau) \right) d\tau
\end{equation}
with the stationary value approximated as $\bar{P}_{gen} \approx \hat{\bar{P}}_{gen}(\textrm{1200s})$. 

\begin{figure}
    \vspace{0pt}
    \centering
   \includegraphics[scale=0.62,trim={0.6cm 0.5cm 0cm 1cm},clip]{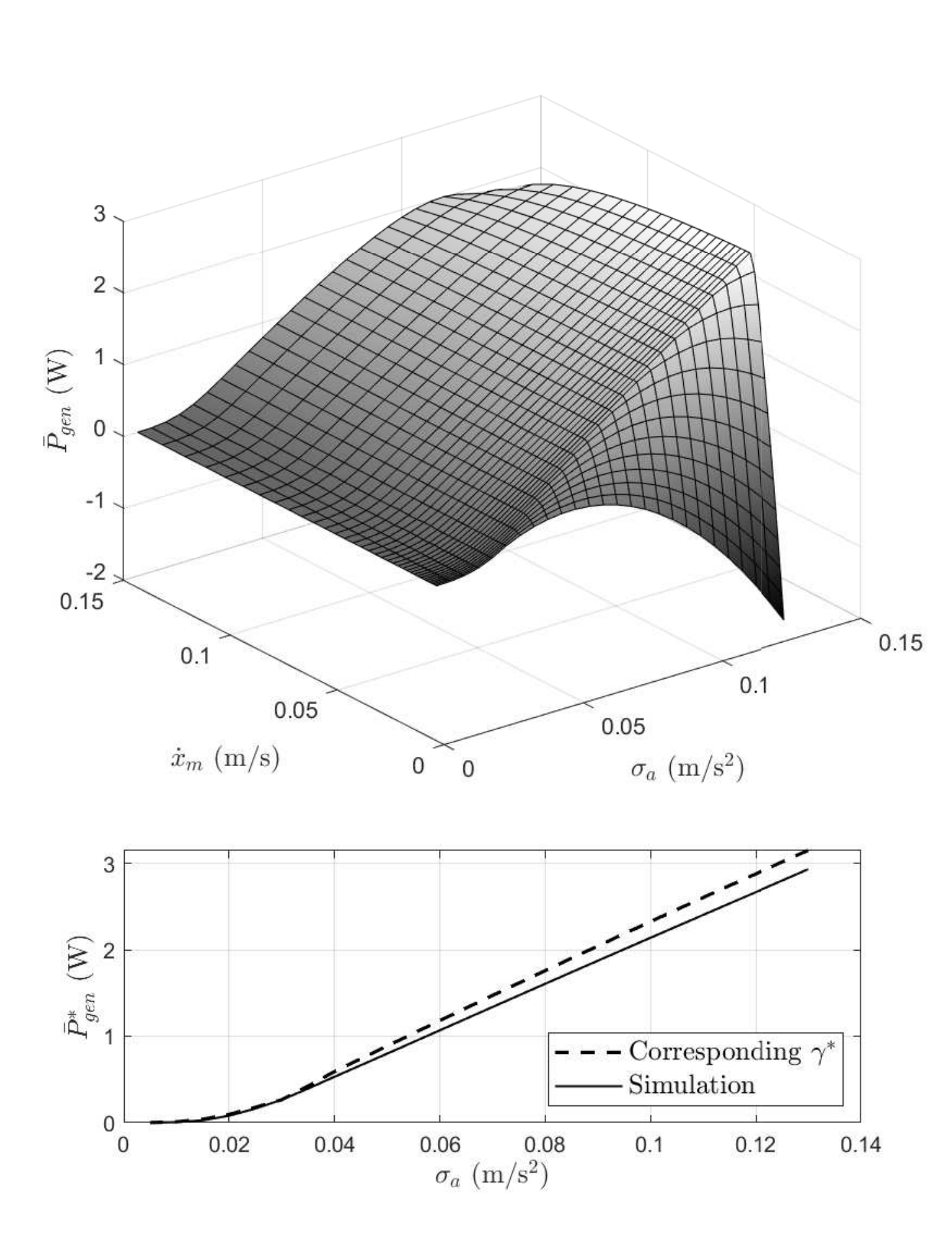}
   \caption{Effect of velocity constraint $\dot{x}_m$ and disturbance intensity $\sigma_a$ on simulated mean generated power (top); and comparison of optimal mean generated power from simulation and corresponding $\gamma$ parameter from the iterative optimization of $\mathcal{K}_q$ (bottom)}
   \label{fig_opt_sim}
   \vspace{-5pt}
\end{figure}

Figure \ref{fig_opt_sim} shows a surface plot of $\bar{P}_{gen}$ corresponding to various $\{\dot{x}_m,\sigma_a\}$ combinations. There are a few trends to note. Clearly, there is a trade-off between velocity regulation and power generation. For very small $\dot{x}_m$, it becomes unnecessary to use field-weakening (i.e., controlling negative $i_d^r$ currents), because constraint \eqref{idq_delta_constraint_2} is more easily satisfied. While this does result in smaller $i_d^{r2}R$ losses, power generation is actually reduced in this case because more of the $i_q^r$ control effort is used to satisfy the velocity constraint \eqref{velocity_con}. In contrast, as $\dot{x}_m$ is made larger, $i_d^{r2}R$ losses increase due to the increased need for field weakening. Also in this case, constraint \eqref{i_q_stochastic} becomes excessively conservative, resulting in $i_q^r$ currents that generate less power. Consequently, there is an optimal tuning of the parameter $\dot{x}_m$, which occurs along the ridge in the $\bar{P}_{gen}$ surface. The optimal $\bar{P}_{gen}^*$ values located along this ridge are plotted versus $\sigma_a$ in the bottom of Figure \ref{fig_opt_sim}. For comparison, we also plot the corresponding optimal $\gamma^*$ parameter from the iterative optimization procedure. We note that $\bar{P}_{gen}^*$ is less than $\gamma^*$, due to the $i_d^{r2}R$ losses and the dynamic saturation of $i_q^r$, which are not accounted for in the optimization algorithm. 
In practice, a look-up table could be used to adapt the $\mathcal{K}_q$ controller according to the disturbance intensity, so that performance remains on the ridge as $\sigma_a$ changes.

\section{Experiment}

The simulation results presented above were experimentally verified via hardware-in-the-loop (HiL) testing. In this section we provide a brief overview of the HiL method and a description of our experimental setup. We then report the HiL results.

\subsection{Overview of HiL}

HiL testing is a cyber-physical experimental method that interfaces numerical models with physical system components in real time. In the civil engineering literature this type of testing is known as real-time hybrid simulation \cite{blakeborough2001development}, and has been used extensively to study the performance of both structural control devices (e.g., \cite{christenson2008large,jiang2013real,chae2013large,chae2014large,friedman2015large}) and vibratory energy harvesting technologies (e.g., \cite{cassidy2011design,asai2021hardware}). To conduct a HiL test, the dynamical system under study is first partitioned into a numerical subsystem (NS) and a physical subsystem (PS). In this research, the PS consists of the electromechanical transducer and power electronics, while the NS is comprised of the linear SDOF oscillator, disturbance filter, and optimized feedback control law.  

A real-time computer (such as a dSpace rapid prototyping system or Speedgoat real-time target machine) is used to simulate the dynamics of the NS. The relative displacement across the subsystem coupling points is applied to the PS using a servo-controlled actuator. The restoring force generated by the PS is measured using a load cell and fed back to the NS via an analog-to-digital converter (ADC) interface, closing the loop. This cycle repeats during each time step of the test. Successful HiL testing requires careful coordination and integration of actuation, sensing, computing, and data acquisition technologies. 

\subsection{HiL testbed} \label{exp_testbed_section}

\begin{figure*}
    \vspace{0pt}
    \centering
   \includegraphics[scale=0.65]{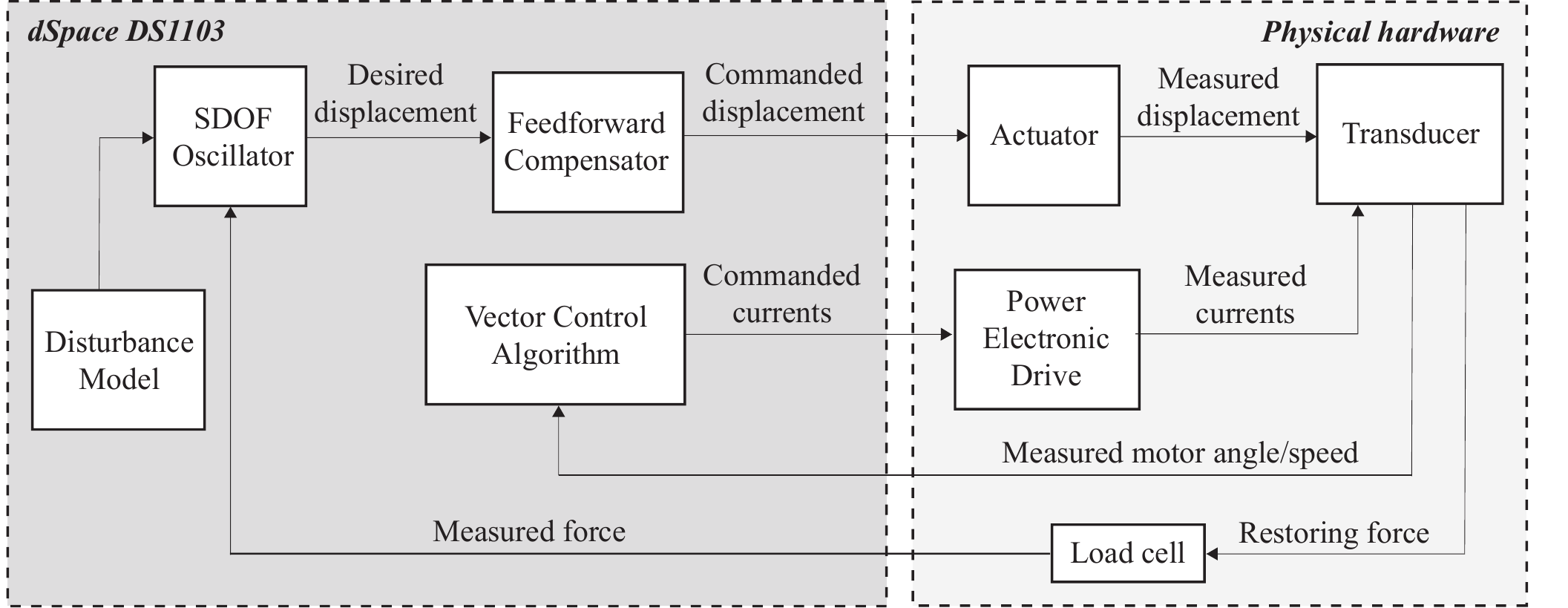}
   \caption{HiL block diagram}
   \label{fig_hil}
   \vspace{-5pt}
\end{figure*}

\begin{figure*}
    \vspace{0pt}
    \centering
   \includegraphics[scale=0.0965]{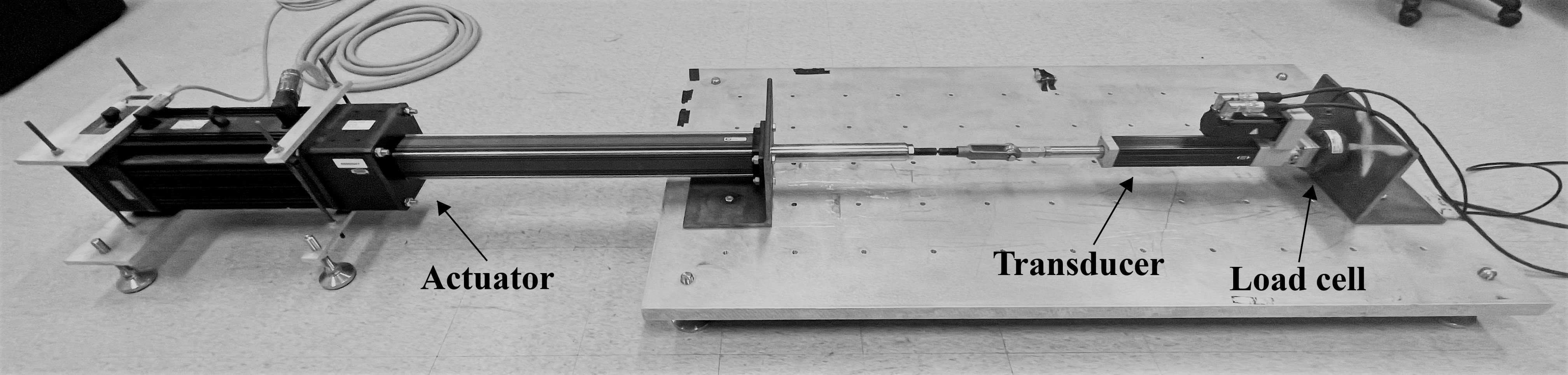}
   \caption{Experimental setup}
   \label{fig_exp}
   \vspace{-10pt}
\end{figure*}

A block diagram of the HiL scheme used in this research is shown in Figure \ref{fig_hil}. The SDOF oscillator and stochastic disturbance models, along with the corresponding vector control algorithm are implemented in Simulink, and simulated on a dSpace DS1103 board in real time at a sampling frequency of 4096 Hz. The physical testbed is shown in Figure \ref{fig_exp}. It is comprised of a 50 cm stroke, 30 kN electromechanical linear actuator, which consists of a Exlar planetary roller screw coupled to a 20 kW Lenze induction motor. The actuator position is controlled using a digitally-programmable Lenze drive, which is interfaced with the dSpace DS1103 unit via the CAN protocol. The drive controller is highly configurable, with nested position, velocity, and current feedback loops, each having tunable gains. In addition, a model-based feed-forward compensator \cite{carrion2007model} is used to further improve the actuator's dynamic response and minimize position-tracking error. 

The transducer used in this study consists of a Kollmorgen AKM24C PMSM, rated at 0.7 kW and 480 V, coupled via ballscrew to a Kollmorgen EC2-series electric cylinder, with a 3.6 kN maximum force rating. Additional transducer data is listed in Table \ref{table:2}. The PMSM is equipped with an internal resolver that provides angular position and velocity measurements. An Analog Devices AD2S1205 resolver-to-digital converter chip is used to interface these measurements with the dSpace DS1103 unit. The transducer is attached to the actuator via a clevis connection, as shown in Figure \ref{fig_exp}. An Interface Model 1210 load cell is used to measure the transducer's restoring force. 

An Agilent N5749A power supply provides the 20 V bus voltage to an Advanced Motion Controls S16A8 PWM servo-drive, which controls the transducer's $i_{abc}$ currents using analog PI feedback loops. The PWM switching frequency of the drive is 33 kHz. The drive also provides measurements of the three-phase currents. The three-phase voltages $v_{abc}$ are measured using a signal conditioning circuit consisting of 11:1 attenuators and active low-pass filters (with approximately 1000 Hz cutoff frequency).

\begin{figure}
    \vspace{0pt}
    \centering
   \includegraphics[scale=0.66,trim={0.4cm 12.8cm 1cm 1.2cm},clip]{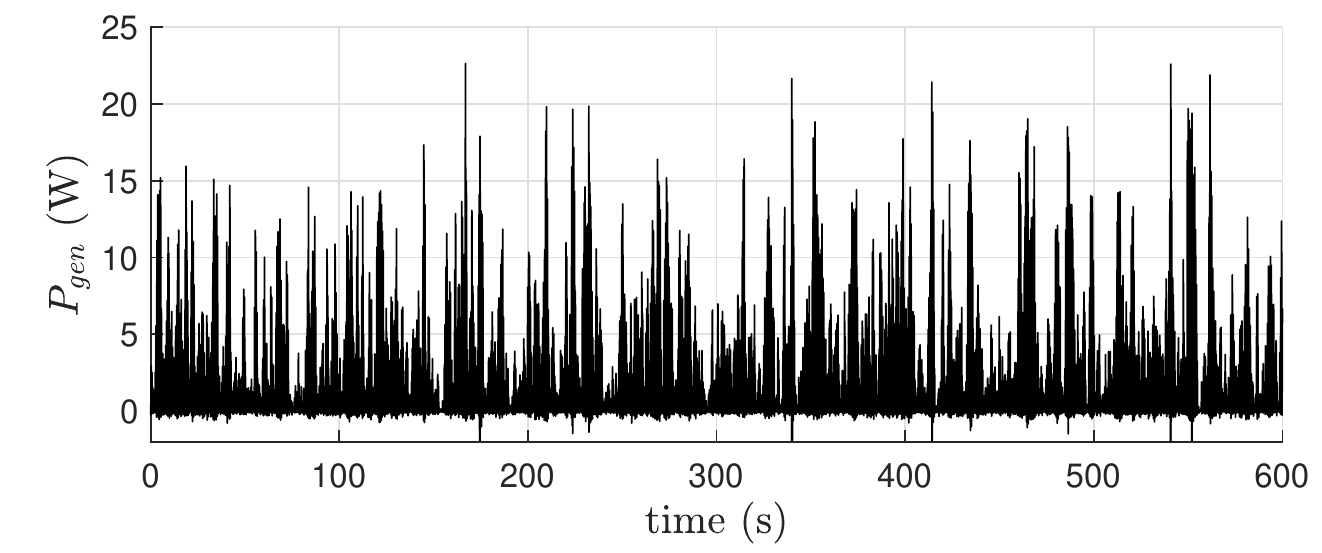}
   \caption{Generated power measured during HiL test with $\sigma_a=0.1~ m/
   s^2$ and $\dot{x}_m=0.0286~ m/s$  }
   \label{fig_typ_p}
   \vspace{-10pt}
\end{figure}

\subsection{Transducer mechanical parameter identification}

Prior to HiL testing, the mechanical transducer model parameters in \eqref{trans_model} were identified using data from a series of characterization experiments. The friction term $f_c$ was determined by back-driving the transducer according to a sinusoidal position profile with a frequency of 0.01 Hz and an amplitude of 25 mm, resulting in extremely low linear velocities and accelerations. It follows that the forces produced during this test could be attributed almost entirely to Coulomb friction. Subsequently, we conducted a 60-second, position sine sweep with frequency content ranging from 0.2-2 Hz and a velocity envelope ranging from 40-70 mm/s. We then used a least squares approach to determine the inertia $J$ and viscous damping $B$ parameters using the data from this second test. 

\subsection{HiL results}

\begin{figure*}
    \vspace{0pt}
    \centering
   \includegraphics[scale=0.53,trim={4cm 0cm 3.4cm 0cm},clip]{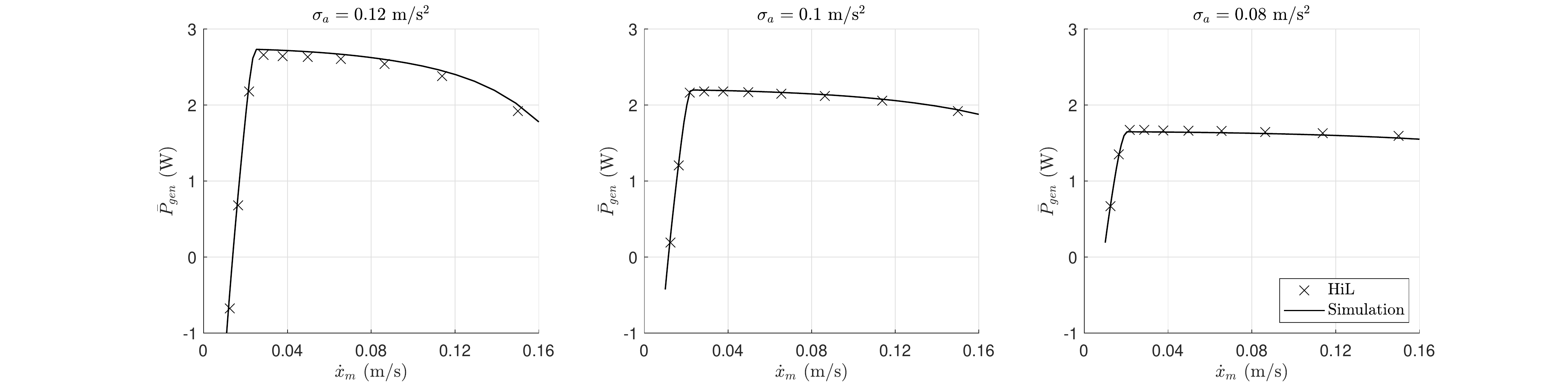}
   \vspace{-5pt}
   \caption{Comparison of mean generated power from HiL tests and simulations for different velocity constraints $\dot{x}_m$ and disturbance intensities $\sigma_a$}
   \label{fig_sigma_data}
   \vspace{-5pt}
\end{figure*}

We conducted a total of 30 ten-minute long HiL tests for different combinations of the $\{\dot{x}_m,\sigma_a\}$ parameters. The experimentally measured $\bar{P}_{gen}$ for each of these cases are plotted in Figure \ref{fig_sigma_data}. Also shown in Figure \ref{fig_sigma_data} are the corresponding ``slices" of the $\bar{P}_{gen}$ surface obtained via simulation shown in Figure \ref{fig_opt_sim}. In general, there is very good agreement between the simulated and experimental results. 

\begin{figure}
    \vspace{0pt}
    \centering
   \includegraphics[scale=0.66,trim={0.4cm 1cm 1cm 1.2cm},clip]{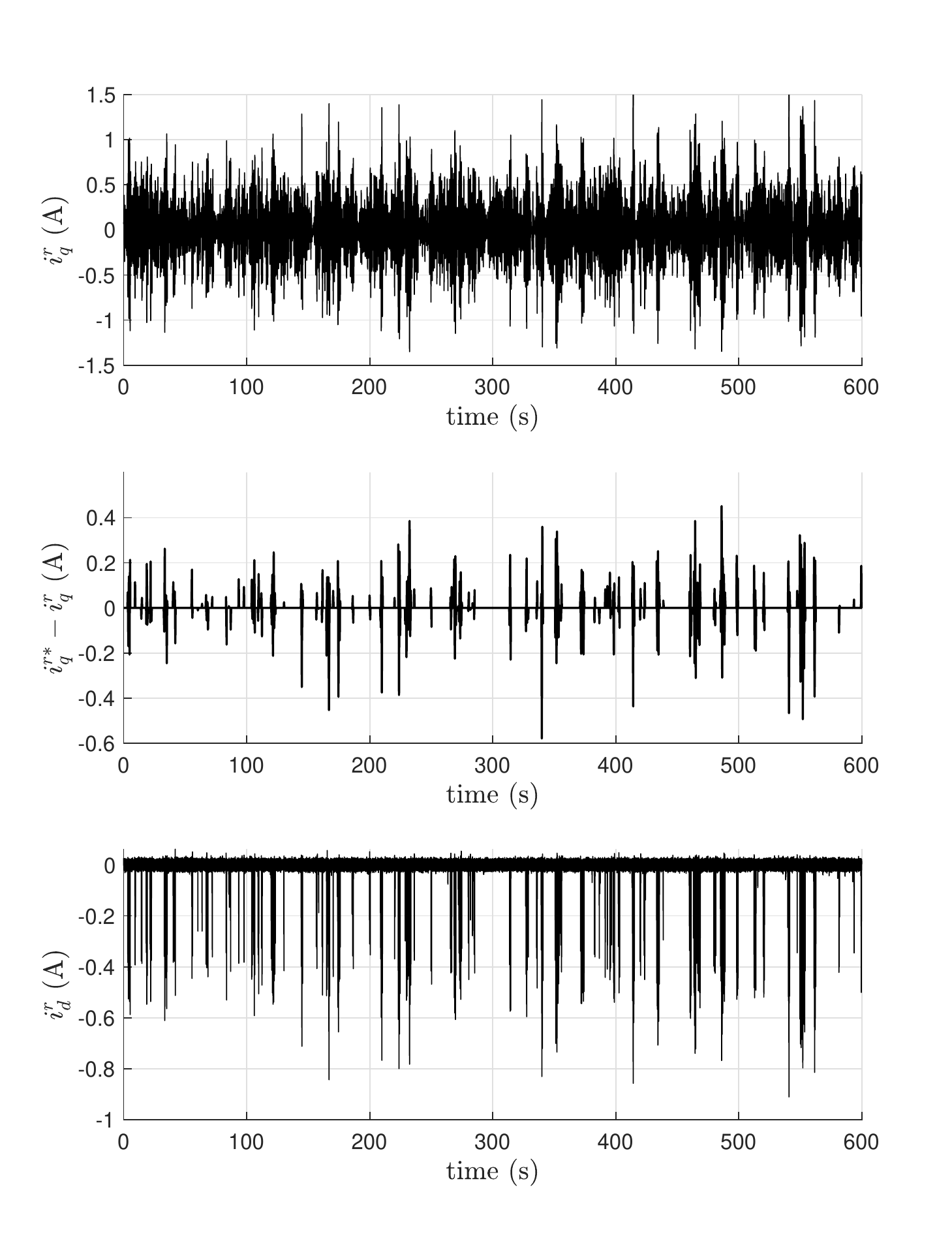}
   \caption{Quadrature-axis current (top); effect of saturation action \eqref{sat_eqn} on quadrature-axis current (middle); and direct-axis current (bottom) measured during HiL test with $\sigma_a=0.1~ m/
   s^2$ and $\dot{x}_m=0.0286~ m/s$  }
   \label{fig_typ_current}
   \vspace{-5pt}
\end{figure}

Figure \ref{fig_typ_p} shows the full generated power time history associated with disturbance intensity $\sigma_a=0.1~ m/s^2$ and velocity constraint $\dot{x}_m=0.0286~ m/s$. The generated power flow was almost entirely unidirectional, validating our assumption used to linearize the efficiency function $h(p(t))\approx 1/\eta$ for the synthesis of $\mathcal{K}_q$. Time histories of the measured $i_q^r$ and $i_d^r$ currents are shown in Figure \ref{fig_typ_current}, as well as the difference between the desired $i_q^{r*}$ produced by $\mathcal{K}_q$ and the dynamically saturated $i_q^r$ from \eqref{sat_eqn}. We see that both saturation and field-weakening are occasionally needed to maintain current feasibility in this case. In addition, the nonzero $i_d^r$ current introduced approximately 0.088 W of resistive power loss, which is 4\% of the mean generated power.

\begin{figure}
    \vspace{0pt}
    \centering
   \includegraphics[scale=0.66,trim={0.4cm 12.8cm 1cm 1cm},clip]{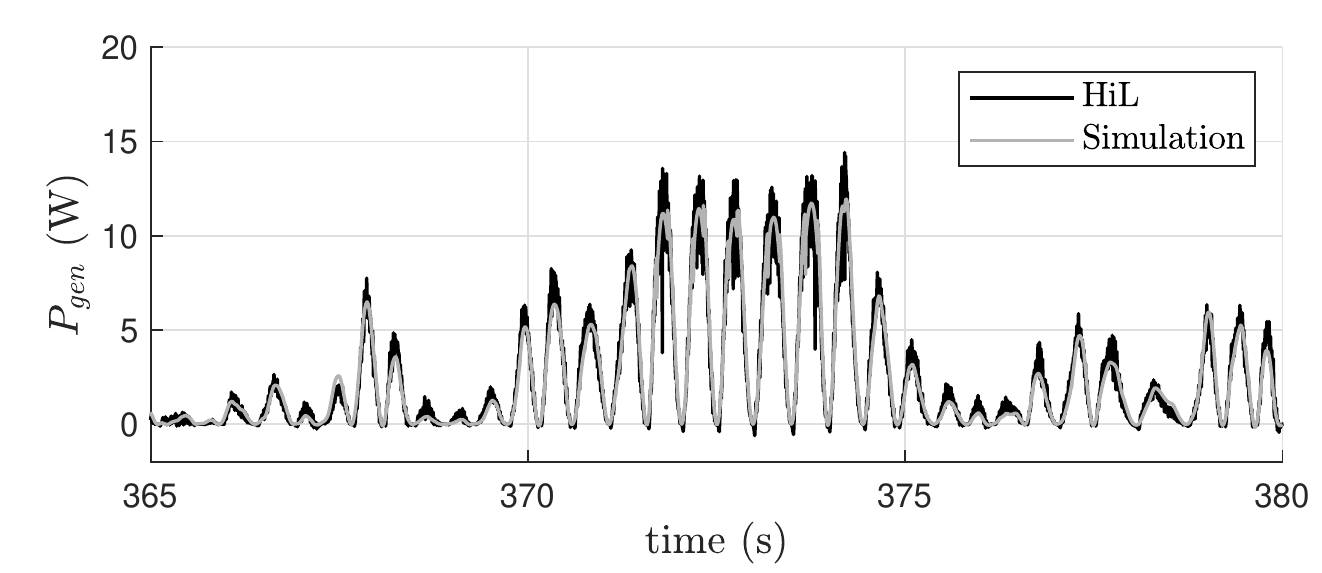}
   \caption{Typical generated power data from HiL test with $\sigma_a=0.1~ m/
   s^2$ and $\dot{x}_m=0.0286~ m/s$ and corresponding numerical simulation }
   \label{fig_typ_p_2}
   \vspace{-5pt}
\end{figure}

\begin{figure}
    \vspace{0pt}
    \centering
   \includegraphics[scale=0.66,trim={0.4cm 1cm 1cm 1cm},clip]{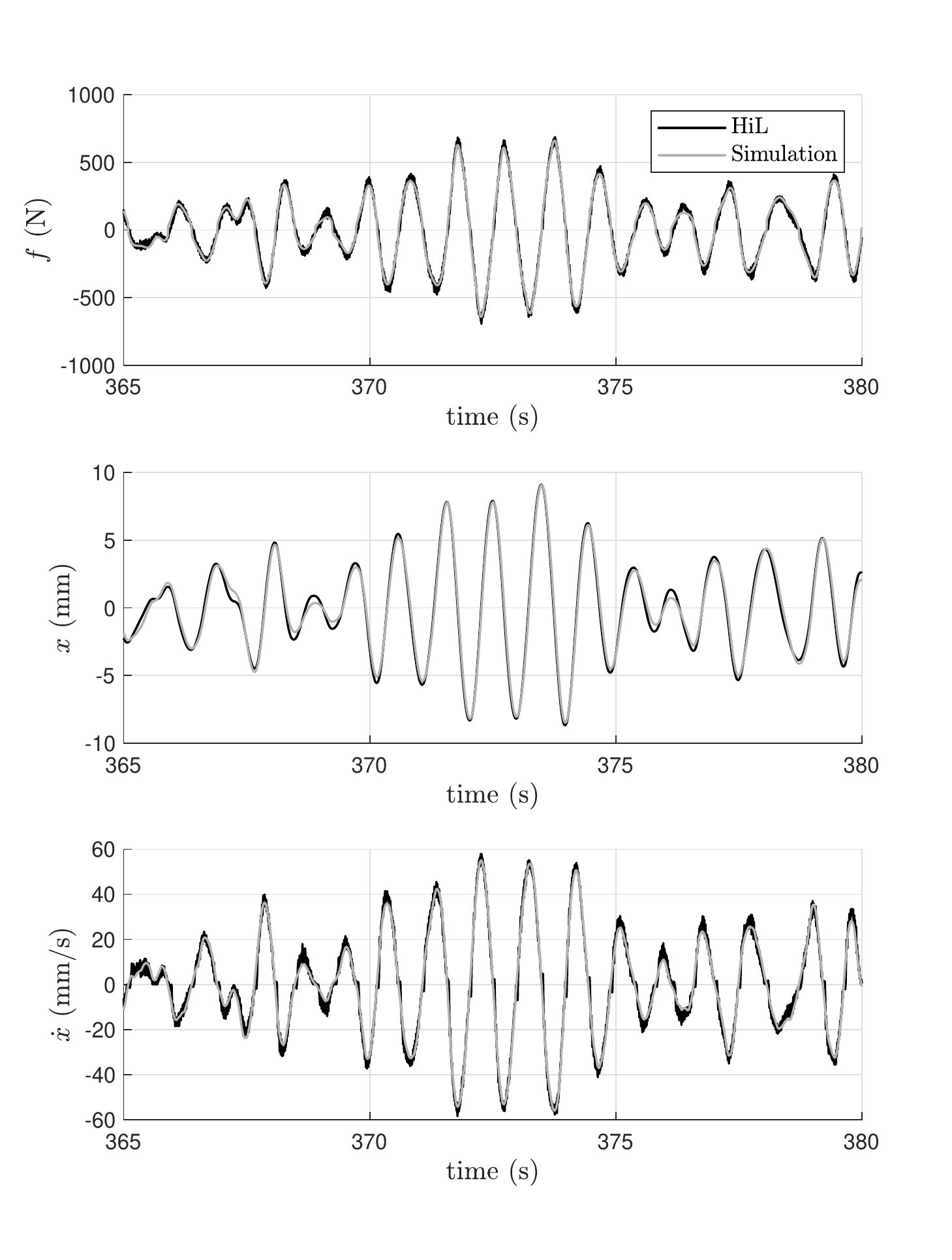}
   \caption{Typical transducer force (top); position (middle); and velocity (bottom) data from HiL test with $\sigma_a=0.1~ m/
   s^2$ and $\dot{x}_m=0.0286~ m/s$ and corresponding numerical simulation}
   \label{fig_typ_data_1}
   \vspace{-5pt}
\end{figure}

\begin{figure}
    \vspace{0pt}
    \centering
   \includegraphics[scale=0.66,trim={0.4cm 1cm 1cm 1cm},clip]{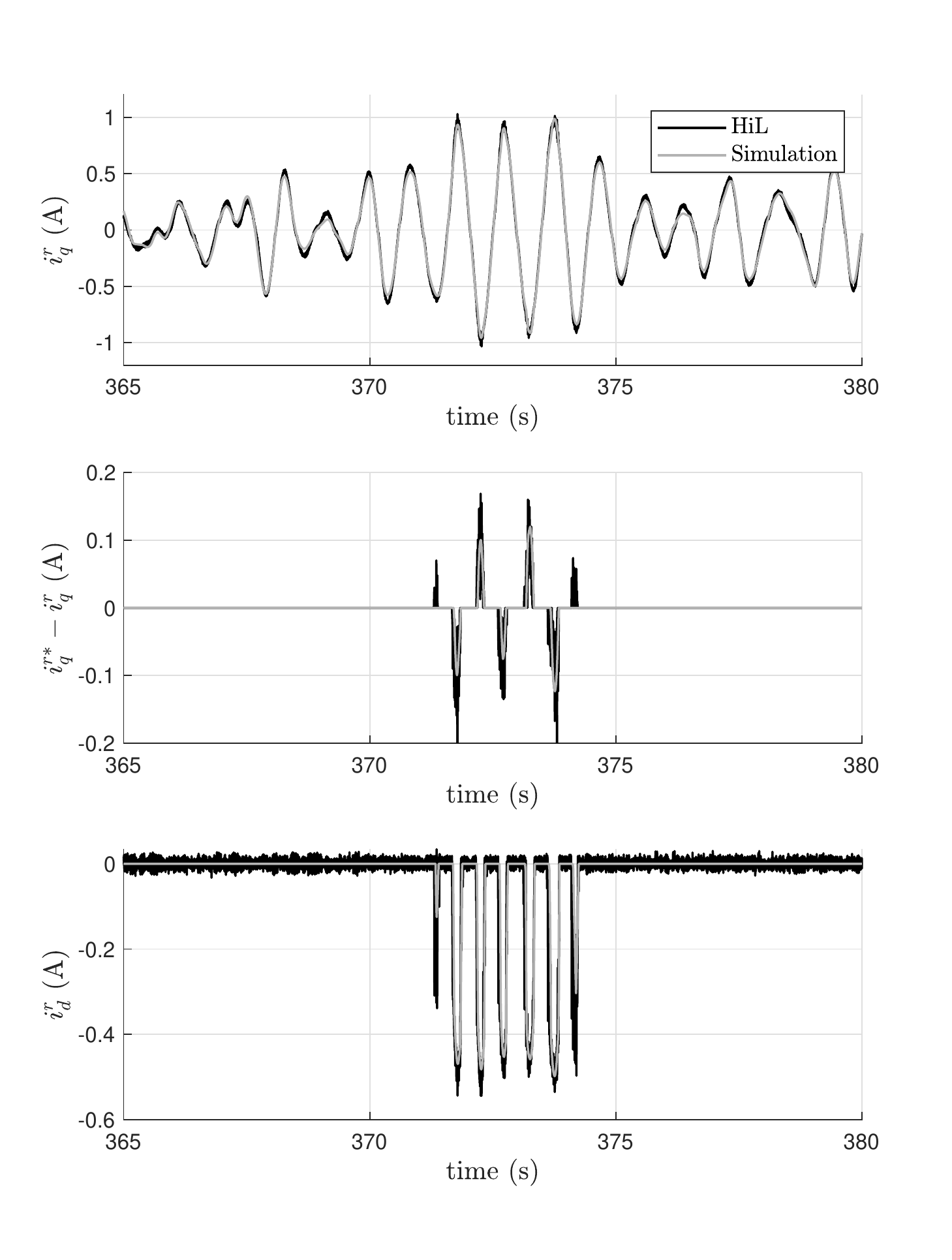}
   \caption{Typical quadrature-axis current (top); saturation action \eqref{sat_eqn} on quadrature-axis current (middle); and direct-axis current (bottom) data from HiL test with $\sigma_a=0.1~ m/
   s^2$ and $\dot{x}_m=0.0286~ m/s$ and corresponding numerical simulation}
   \label{fig_typ_data_2}
   \vspace{-10pt}
\end{figure}

Finally, Figures \ref{fig_typ_p_2}-\ref{fig_typ_data_2} contain a variety of data comparing the HiL and simulation results over a shorter time-span of 15 seconds. There was consistently good agreement between all measured signals, again confirming the validity of using \eqref{trans_model} to model the transducer's mechanical dynamics. One important thing to note is that dynamic saturation of $i_q^r$ and negative $i_d^r$ occur when the transducer velocity is large, as expected given constraint \eqref{idq_delta_constraint_2}.


\section{Conclusions}

In this paper, we presented a technique to design feedback control laws that approximately maximize the power generation of a three-phase energy harvester. We assumed a stochastic vibratory disturbance model, and utilized a vector control framework. While our proposed synthesis method is heuristic, it explicitly accounts for constraints imposed on the harvester's currents due to a finite power bus voltage. We first designed the rotor reference frame quadrature-axis current controller $\mathcal{K}_q$ via an iterative multi-objective optimization procedure using a linearized transducer model. Subsequently, we implemented field-weakening via the direct-axis current controller $\mathcal{K}_d$. The multi-objective optimization imposed competing, mean-square constraints on the $i_q^r$ current and transducer velocity $\dot{x}$. Through a simulation example, we determined that there existed an optimal tuning of the velocity constraint parameter $\dot{x}_m$, which produced the highest mean generated power for a given disturbance intensity. The simulation results were then confirmed experimentally via hardware-in-the-loop testing of an actual PMSM transducer. Finally, we emphasize that the methodology developed herein is suboptimal and future work should focus on the parallel, rather than sequential, design of $\mathcal{K}_q$ and $\mathcal{K}_d$ to both maximize $\bar{P}_{gen}$ and ensure current feasibility. 

\bibliographystyle{IEEE}   
\bibliography{vector_control}   

\end{document}